\documentclass[onecolumn,tightenlines,nofootinbib,preprintnumbers,amsmath,amssymb]{revtex4}
\usepackage[usenames,dvipsnames]{color}
\usepackage{caption2}
\usepackage{dcolumn}
\usepackage{makecell}
\usepackage{graphicx}
\usepackage{subfigure}
\usepackage{epstopdf}
\usepackage{bm}
\usepackage{lscape}
\usepackage{slashed}
\usepackage{booktabs}
\renewcommand{\arraystretch}{0.9}
\usepackage{setspace}

\begin{document}
	\begin{spacing}{1.5}
		\title{CP asymmetry from the effect of the isospin symmetry breaking  during B-meson decay}
		\author{Wan-Ying Yao$^{1}$\footnote{Email: 1904132420@qq.com}, Gang L\"{u}$^{1}$\footnote{Corresponding author Email: ganglv66@sina.com}, 	Xin-Heng Guo $^{2}$\footnote{Email: xhguo@bnu.edu.cn}, Hai-Feng Ou$^{1}$\footnote{Email: ouhaifeng@haut.edu.cn}}
		\affiliation{\small $^{1}$ Institute of Theoretical Physics, College of Physics, Henan University of Technology, Zhengzhou 450001, China\\
		\small $^{2}$ College of Nuclear Science and Technology, Beijing Normal University, Beijing 100875, China\\}
	\begin{abstract}
		
		The direct CP asymmetry in quasi-two-body decays of $B \rightarrow (V\rightarrow  \pi^{+}\pi^{-})P $ is investigated in the perturbative QCD method, where P represents a pseudoscalar meson and V refers to $\rho$, $\omega$ and $\phi$ mesons, respectively.				
		We present the amplitude of the quasi-two-body decay process and investigate the effects of mixed resonances involving $\rho^{0}-\omega$, $\rho^{0}-\phi$ and $\omega-\phi$, while considering the impact of isospin symmetry breaking. 	We observe a significant CP asymmetry when the invariant mass of the $\pi^{+}\pi^{-}$ pair is within the resonance ranges of  $\rho$, $\omega$ and $\phi$ mesons. Consequently, we proceed to quantify the regional CP asymmetry in these resonance regions. A significant difference is observed when comparing results obtained with and without interferences of the three vector mesons and isospin conservation.
		The CP asymmetry results obtained from the three-body decay process, without interference due to isospin conservation by the perturbative QCD method, are in agreement with the newly updated data acquired by the LHCb experiment.
		
	\end{abstract}
	\maketitle
	
	\section{\label{intro}Introduction}
	
	CP asymmetry, which plays a crucial role in elucidating the matter-antimatter asymmetry of the Universe, holds significant theoretical and experimental significance in the field of Particle physics \cite{Shaposhnikov:1991cu,Okun:1967cxa}.	
	CP asymmetry has been experimentally observed in the K, B and D meson decay process		
	\cite{KTeV:1999aiu,BaBar:2001oxa,Belle:2001zzw}. CP asymmetry is closely related to weak phases and strong phases. The CKM matrix describes the mixing between different quark flavors and carries the phase information of CP asymmetry. The weak phase originates from the CKM matrix, which represents the transformation of quarks from their mass eigenstates to weak interaction eigenstates. Meanwhile, the strong phase arises from hadronic matrices and intermediate states \cite{Cabibbo:1963yz}. Recently, CP asymmetry has been observed in the course of the B-meson multi-body decay in experiment \cite{LHCb:2019jta,LHCb:2022tuk}.		 
	The investigation of CP asymmetry in multi-body decays of B mesons plays an increasingly pivotal role in both scrutinizing the CKM mechanism of the SM and exploring novel sources of CP asymmetry \cite{Kobayashi:1973fv}.

	In contrast to the two-body decay, the three-body decay of B meson encompasses both resonance and non-resonance contributions, as well as being influenced by final state re-scattering in $KK\rightarrow\pi\pi$ interactions \cite{LHCb:2013ptu}.
	The observation of significant CP asymmetry in localized regions of phase space for charmless three-body B-meson decays has been reported through model-independent analyses. In the LHCb experiment, researchers extensively investigated the complete phase space of B meson decay, focusing on a specific invariant mass region. The information on CP asymmetry in the $B^\pm \rightarrow  \pi^{+}\pi^{-}\pi^\pm$ decay process is represented by $A_{CP}^{inte}$ in the full phase space, with the measured data being  $A_{CP}^{inte}=5.8\pm0.8\pm0.9\pm0.7\%$. In the specified invariant mass region, for the low invariant mass region with $m^2_{\pi^+\pi^- \rm ~low}<0.4~{\rm GeV}^2$ and high invariant mass region with $m^2_{\pi^{+}\pi^{-}\rm~high}<15~{\rm GeV}^2$, the expression for CP asymmetry is $A_{CP}^{low}$, with the corresponding CP asymmetry value being $A_{CP}^{low}=58.4\pm8.2\pm2.7\pm0.7\%$. While in the range of rescattering with $m^2_{\pi^{+}\pi^{-}\rm~high}<15~{\rm GeV}^2$, the expression for CP asymmetry is $A_{CP}^{resc}$, with the corresponding CP asymmetry value being $A_{CP}^{resc}=17.2\pm2.1\pm1.2\pm0.7\%$ \cite{LHCb:2014mir,LHCb:2013lcl}.	 			 
	The data is extracted from the $B^{\pm} \rightarrow \pi^{+} \pi^{-} K^{\pm}$ decay for the integrated CP asymmetry in full phase space, where $A_{CP}^{inte}$ is measured to be $2.5\pm0.4\pm0.4\pm0.7 \%$. Additionally, measurements are performed in specific regions: the low invariant mass region ($0.08~{\rm GeV}^2 < m_{\pi^+\pi^- ~low}^2 < 0.66~{\rm GeV}^2$) and the high invariant mass region ($m_{\pi^\pm K^\pm ~high}^2 < 15~{\rm GeV}^2$), yielding $A_{CP}^{low}$ of $67.8\pm7.8\pm3.2\pm0.7 \%$ and measurements in rescattering regions with $1.0~{\rm GeV}<m_{\pi^+\pi^-}<1.5~{\rm GeV}$ resulting in $A_{CP}^{resc}$ of $12.1\pm1.2\pm1.7\pm0.7 \%$ \cite{LHCb:2013ptu, LHCb:2014mir}.

The three-body decay process has complex dynamical information. 
Based on the available experimental data, there is a strong indication of significant CP symmetry within the special
region.   
The decay of three-body B mesons was analyzed by using the Dalitz diagram in the LHCb experiment 
 \cite{LHCb:2013ptu,LHCb:2014mir}.  
 The characteristics of the resonance amplitudes can be clearly observed from the Dalitz diagram, revealing that scalar and vector resonances with invariant masses less than approximately $1 GeV/c^{2}$ are dominant 
  \cite{Cheng:2016shb}. 		  
In the decay of charged B meson, it is clearly observed from the Dalitz diagram that there are obviously CP asymmetry  when the invariant masses of $ \pi^{+}\pi^{-}$ are in the $ \rho^{0 }(770)$ and $f_{0 } (980)$ resonance regions \cite{Qi:2018lxy,Qi:2023pls}. Until recently, a resonance model that accurately describes these effects has been lacking \cite{LHCb:2019jta,LHCb:2022tuk}. Our previous results show that  there is a large CP asymmetry at
the regions of
 the $ \rho^{0 }(770)$ and $\omega (782)$ resonance
 for B meson decay process \cite{Yuan:2023ujm,Huber:2016xod,Shi:2022ggo,6Lu2017, Lu:2023yxa}.

	The charmless decay of B mesons to three hadrons is predominantly governed by a quasi-two-body process involving an intermediate resonance state, the effects of resonance can be accurately described using the Breit-Wigner (BW) formalism. 
	We aim to investigate the CP asymmetry of vector meson resonance in the three-body decay of B meson using the PQCD method. With a comprehensive exploration of CP asymmetry, various approaches have been employed to describe the multibody decays of B mesons, including QCD factorization (QCDF) \cite{Yuan:2023ujm,Huber:2016xod} and PQCD \cite{Shi:2022ggo,Chang:2024qxl}. 
	In the decay process of the B meson, different methods  introduce different phases, thereby affecting the value of CP asymmetry.	
	In the QCDF method, the b-quark mass $m_b$ is taken to approach infinity in the decay of the B meson, while neglecting the higher-order contribution of $1/m_b$.	 
	 The consideration is solely focused on the longitudinal momentum, while the transverse momentum, being relatively small in magnitude, can be disregarded when compared to its longitudinal counterpart	 
	 \cite{Beneke:2021rjf}. The PQCD approach preserves the $k_T$ transverse momentum, selects an appropriate scale and introduces Sudakov factors to resolve endpoint divergences \cite{Hua:2020usv,Yang:2022ebu}. 

	According to the Vector Meson Dominated (VMD) model, the polarization of photons in vacuum leads to the emergence of vector particles such as $\rho^{0}$, $\omega$ and $\phi$. The annihilation of $e^{+}e^{-}$ into photons results in the production of a pair of $\pi^{+}\pi^{-}$ through an intermediate neutral vector meson. 	
	The decay of $\rho^{0}$ into $\pi^{+}\pi^{-}$ is an isospin-conserving process, while the decays of $\omega$ and $\phi$ into pair of $\pi^{+}\pi^{-}$ involve isospin asymmetry. To better describe these decay processes, we utilize a unitary matrix transformation to convert the isospin of the intermediate state (non-physical states) into a physical field. By observing interference between the resonant mesons ($\rho^{0}$, $\omega$ and $\phi$), valuable insights into their kinetic mechanisms can be obtained. Furthermore, it should be noted that the presence of intermediate resonance hadrons contributes to the formation of a new strong phase, which may have implications for CP asymmetry in hadronic decays.  			 
	We investigate the impact of mixed resonance of intermediate particles on CP asymmetry in the three-body decay process of B meson. By examining the decay of vector meson resonances, it will enable more precise measurement of CP asymmetry in future experiment. Additionally, we also discusses regional variations in CP asymmetry for comparison with experimental results.

	The layout of the present paper is as follows: In Sec. II we discuss resonant contributions to three-body  decays of B meson. 
	The quasi-two-body method $B\to VP$( $V\to PP$) is employed to investigate the CP asymmetry. 
	In Sec. III  
	is dedicated to the examination of direct CP asymmetry.  
	The regional CP asymmetries arising from the resonance effects of $\rho^{0}$, $\omega$ and $\phi$ are taken into consideration. 
	We illustrate the form of the three-body decay amplitude of B meson  after considering above resonance effect and present the CP asymmetry. The numerical results and discussion on the local CP asymmetry is presented in Sec. IV. 
	Sec. V contains our conclusions.

	\section{\label{sum}THE EFFECT OF VECTOR  MESON RESONANCE  ON THREE-BODY  DECAY OF B MESON}
	\subsection{\label{subsec:form} mixing mechanism}

	Based on the VMD model, electrons and positrons annihilate to generate photons, which are further polarized to form vector meson  $\rho^{0}$, $\omega$, $\phi$. Then, these vector mesons decay into pair of $\pi^{+}$$\pi^{-}$ mesons \cite{6Lu2017}. In this mechanism, the mixing parameters corresponding to two vector particles can be obtained by using the electromagnetic form factor of the $\pi$ meson.

	Since the $\rho^{0}_{I}$ ($\omega_{I}$, $\phi_{I}$) resonance state is a non-physical state, we use the  unitary matrix R to transform the isospin field into the physical field: $\rho_{I}^{0}$ ($\omega_{I}$,$\phi_{I}$ ) $\rightarrow \ \rho^{0}$ ($\omega$, $\phi$). The diagonal elements of the unitary matrix R are equal to 1, while the off-diagonal elements encode information regarding the $\rho^{0}-\omega-\phi$ mixture. The contribution of high order terms is ignored in the process of this transformation. 	 	  
	The resonance effects for $A_{\rho\omega}, B_{\omega\phi}$ and $C_{\rho\phi}$ are defined by a set of mixed amplitude parameters, where it holds that $A_{\rho^{0}\omega}=-A_{\omega\rho^{0}}$, $B_{\omega\phi}=-B_{\phi \omega}$, and $C_{ \rho^{0} \phi } = - C _{ \phi \rho^{0} }$. 	 	 	 	  
	The amplitude represents a first-order approximation that is dependent on the variable s, which in turn is related to the square of momentum. The expression  is

\begin{equation}
	\left (
	\begin{array}{lllll}
		\rho^0\\[0.5cm]
		\omega\\[0.5cm]
		\phi
	\end{array}
	\right )
	=
	R(s)
	\left (
	\begin{array}{lll}
		\rho^0_I\\[0.5cm]
		\omega_I\\[0.5cm]
		\phi_I
	\end{array}
	\right )
	=	\left (
	\begin{array}{lll}
		~~~~1 &\hspace{0.5cm} -A_{\rho\omega}(s) &
		\hspace{0.5cm}  -C_{\rho\phi}(s)\\[0.5cm]
		\displaystyle  A_{\rho\omega}(s) &  \hspace{0.5cm}~~~ 1 &
		\hspace{0.5cm}  -B_{\omega\phi}(s)  \\[0.5cm]
		\displaystyle  C_{\rho\phi}(s)
		& \hspace{0.5cm}  \displaystyle  B_{\omega\phi}(s)& \hspace{1.0cm} 1
	\end{array}
	\right )\left (
	\begin{array}{lll}
		\rho^0_I\\[0.5cm]
		\omega_I\\[0.5cm]
		\phi_I.
			\end{array}
		\right )		
	\label{L1}.
	\end{equation}
	\noindent

	According to the transformation relation of the R matrix, the expression form of the resonance state in the physical state is given: $ \phi=C_{\rho^{0}\phi  }(s) \rho_{I}^{0}+B_{\omega \phi}(s) \omega_{I}+\phi_{I}$, 
	$\omega=A_{ \rho^{0}\omega }(s) \rho_{I}^{0}+\omega_{I} -B_{\omega \phi}(s) \phi_{I}$, 
	$\rho^{0}=\rho_{I}^{0}-A_{\rho^{0}\omega }(s) \omega_{I}-C_{\rho^{0}\phi }(s) \phi_{I}$ \cite{Lu:2023yxa}.

   We define the mixing parameters $\Pi_{V_{i} V_{j}}$  ($V_i$ and $ V_j$ represent distinct vector mesons) as \cite{Shi:2022ggo,2lu2022}:
	\begin{equation}
		\begin{split}	
		&A_{\rho^0 \omega}=-A_{\omega\rho^0}=\frac{\Pi_{\rho^0 \omega}}{({s-m_{\rho^0}^{2}+im_{\rho^0} \varGamma_{\rho^0}})-({s-m_{\omega}^{2}+im_{\omega} \varGamma_{\omega}})}	 \\
			&B_{\omega \phi}=- B_{\phi\omega}=\frac{\Pi_{\omega \phi}}{({s-m_{\omega}^{2}+im_{\omega} \varGamma_{\omega}})-({s-m_{\phi}^{2}+im_{\phi} \varGamma_{\phi}})} \\
			&C_{\rho^0 \phi}=- C_{\phi\rho^0}=\frac{\Pi_{\rho^0 \phi}}{({s-m_{\rho^0}^{2}+im_{\rho^0} \varGamma_{\rho^0}})-({s-m_{\phi}^{2}+im_{\phi} \varGamma_{\phi}})}.
		\end{split} 
	\end{equation}
	
	The ${s-m_{v}^{2}+im_{v} \varGamma_{v}}$ is the reciprocal of the vector meson propagator \cite{AG}, where $s$ represents the invariant mass square of pair of $\pi^+\pi^-$ mesons. 
The decay width of the vector meson is denoted by $\varGamma_{V}$. The approximation for the decay width of $\rho^0$ is given as $\varGamma_{\rho^0} \rightarrow \pi \pi$. The mass of the vector meson is represented by $m_{V}$.

	The parameters $A_{\rho^0 \omega}$, $	B_{\omega \phi}$, $C_{\rho^0 \phi}$ and ${\Pi_{V_iV_j}}$ are considered as first-order approximations. Any product of two parameters can be disregarded due to its higher-order nature. We hereby define: 
	\begin{equation}
		\begin{split}
		&\widetilde{\Pi}_{\rho^0\omega}=\frac{({s-m_{\rho^0}^{2}+im_{\rho^0} \varGamma_{\rho^0}})\Pi_{\rho^0\omega}}{({s-m_{\rho^0}^{2}+im_{\rho^0} \varGamma_{\rho^0}})-({s-m_{\omega}^{2}+im_{\omega} \varGamma_{\omega}})}  
		\\
	&\widetilde{\Pi}_{\rho^0\phi}=\frac{({s-m_{\rho^0}^{2}+im_{\rho^0} \varGamma_{\rho^0}})\Pi_{\rho^0\phi}}{({s-m_{\rho^0}^{2}+im_{\rho^0} \varGamma_{\rho^0}})-({s-m_{\phi}^{2}+im_{\phi} \varGamma_{\phi}})}   
		 \\
	&\widetilde{\Pi}_{\omega\phi}=\frac{({s-m_{\phi}^{2}+im_{\phi} \varGamma_{\phi}})\Pi_{\omega\phi}}{({s-m_{\omega}^{2}+im_{\omega} \varGamma_{\omega}})-({s-m_{\phi}^{2}+im_{\phi} \varGamma_{\phi}})}  .
			\end{split}
	\end{equation}

	Through the derivation process of the mixed parameters, it can be inferred that a correlation exists between the mixed parameters and $s$. These mixed parameters amalgamate the contributions from both resonant and non-resonant components arising due to isospin symmetry breaking effects in the direct decay process, while also exhibiting momentum dependence. 
 	Based on the available experimental results, we are able to accurately determine the mixing parameters $\Pi_{\rho^0 \omega }=-4470 \pm 250 \pm 160-i(5800 \pm 2000 \pm 1100)  \mathrm{MeV}^{2}$ in the vicinity of the $\rho^0$ meson. The mixing parameters $\Pi_{\phi\rho^0}=720 \pm 180 -i(870 \pm 320) \mathrm{MeV}^{2}$ were obtained near the $\phi$ meson \cite{CE2009,MN2000}. 
	Isospin symmetry is manifested in the decay of B meson as the $\rho$ meson decay to  $\pi ^+\pi ^-$ satisfies isospin conservation, while the decay of $\omega$ and $\phi$  mesons violates this conservation.
	Hence, we do not take into account  the mixing parameter $\widetilde{\Pi}_{\omega\phi}$, which presents the influence of $\omega$ meson and $\phi$ meson mixing resonance.

	\subsection{\label{subsec:form} CP asymmetry induced by interference}

	The decay {$B\rightarrow (V\rightarrow \pi ^+\pi ^-) \pi(K)$ } is modeled as a quasi-two-body process, where V represents an intermediate vector meson. 
	The presence of vector meson resonances in the decay process significantly impacts the strong phase, tree and penguin amplitudes.  
	In Fig. 1, for the decay of $B\rightarrow [\rho^0(\omega,\phi) \rightarrow \pi ^+\pi ^-]\pi(K)$, we investigate the impacts of various vector meson resonances. 
	
	\begin{figure}[htbp]
		\centering
		\subfigure{\includegraphics[height=11cm,width=18cm]{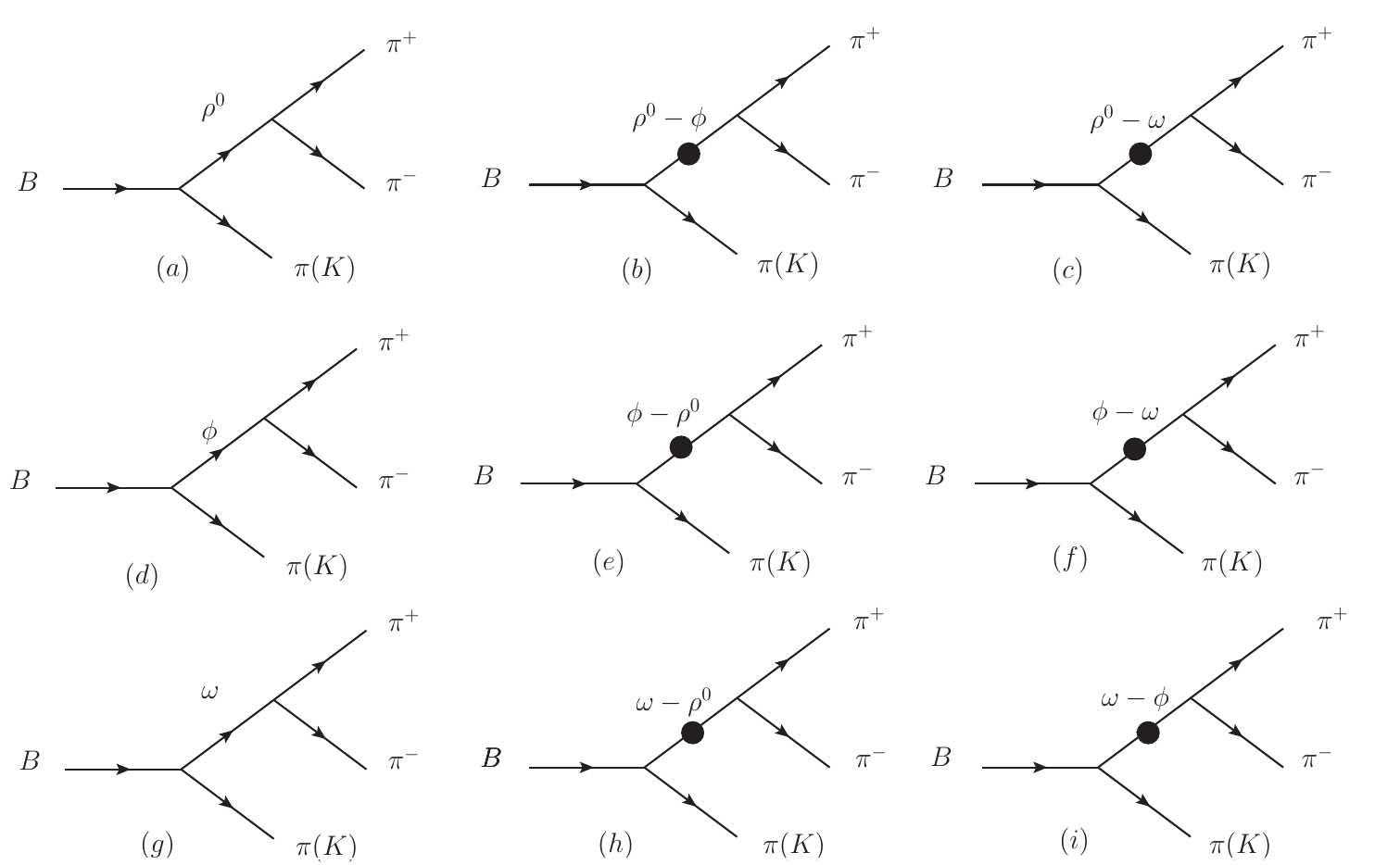}}
		\\
		\caption{The diagrams of ${B}\rightarrow\pi ^+\pi ^-\pi(K)$ decay process.}
		\label{fig2}
	\end{figure}

	According to the three-vector mesons mixing mechanism and neglecting the contribution of higher order terms, we can derive the decay diagram for the $B\rightarrow [\rho^0(\omega,\phi) \rightarrow \pi ^+\pi ^-]\pi(K)$ decay process, as illustrated in diagrams $\left( a \right) \thicksim \left(i \right)$ of Fig. 1.   
	In diagram (a), it can be observed that the ${B}$ meson directly decays to produce a pair of $\pi ^+\pi ^-$ mesons through the intermediate state of $\rho^0$ meson. Simultaneously, it is evident that both the $\omega$ and $\phi$ mesons also contribute to the production of pair of $\pi ^+\pi ^-$ mesons through direct decays: $\phi \rightarrow \pi \pi $  and  $\omega \rightarrow \pi\pi $, as shown in diagrams (d) and (g), respectively.

	We illustrate various effects arising from mixed resonances in diagram (b),(c),(e),(f),(h) and (i) of Fig. 1.
	In addition to the direct decay of vector mesons ($\rho^0$, $\omega$, $\phi$) into a pair of $\pi^+\pi^-$ mesons, the production of such meson pairs can also occur through resonances of vector mesons. During this decay process, isospin symmetry is violated. The black dots in the diagram represent mixing parameters expressed as ${\Pi}_{\mathrm{{V_iV_j}}}$. As shown in diagram (e) and diagram (h), the $\rho ^0$ meson decays into a pair of $\pi ^+\pi ^- $ mesons through the  $\phi$ and $\omega$ resonances, respectively. The corresponding mixing parameters are ${\Pi}_{\mathrm{\rho^0\phi}}$ and ${\Pi}_{\mathrm{\rho^0\omega}}$, where ${\Pi}_{\mathrm{\rho^0\phi}}$ is generated in the $\phi$ resonance process and ${\Pi}_{\mathrm{\rho^0\omega}}$ is generated in the $\omega$ resonance process. Diagram (b) and diagram (c) show the processes of the $\phi$ and $\omega$ mesons decaying into a pair of $\pi ^+\pi ^- $ mesons through $\rho^0$ resonances. The corresponding mixing parameters are ${\Pi}_{\mathrm{\phi\rho^0}}$ and ${\Pi}_{\mathrm{\omega\rho^0}}$.

The decay of the $\omega$ and $\phi$ mesons into a pair of $\pi^+\pi^-$ mesons violates the law of isospin conservation. However, considering the small mixing parameter, the contribution arising from the mixing of $\omega$ and $\phi$ can be treated as a higher-order term that may be neglected. Consequently, we can disregard the contributions from Diagrams (b), (c), (f) and (i) since they correspond to higher-order terms. Additionally, any high-order contribution resulting from multiplying two parameters can also be disregarded; thus cases involving multiple mixed parameters are excluded. 
 An amplitude analysis of this decay channel reveals that $\rho^0$ dominates the contribution  \cite{Pham:2006pq,Belle:2002ezq,CLEO:2000xjz}. However, it is crucial to consider the strong phase generated by the mixed resonance and its impact on CP asymmetry. It should be noted that the mixed parameters ${\Pi}_{\mathrm{\rho^0\phi}}$ and ${\Pi}_{\mathrm{\rho^0\omega}}$ are approximations at first order.

The corresponding amplitude expressions for diagram (a), diagram (e), and diagram (h) in Fig. 1 are denoted as $A_{a}$, $A_{e}$, and $A_{h}$. Herein, we present the amplitude expression for the three-body decay process $B\rightarrow\pi ^+\pi ^-\pi(K)$ as follows:
	\begin{eqnarray}
		A=A_{a}+A_{e}+A_{h}.
	\end{eqnarray}

According to Fig. 1, the present paper employs the quasi-two-body method to calculate the CP asymmetry of B meson. ${\cal A}(\rho ^0{\pi})$ represent the decay amplitudes of the $B\rightarrow\rho^0\pi$ decay processes. ${s-m_{\rho^0}^{2}+im_{\rho^0} \varGamma _{\rho^0}}$ is the reciprocal of the $\rho^0$ meson propagator, ${g^{\rho ^0\rightarrow \pi ^+\pi ^- }}$ represents the coupling constant of $\rho \rightarrow \pi \pi $  \cite{LW1964}.	
In diagram (a), the expression for the amplitudes of the direct decay of the $\rho^0$ meson is shown:		
\begin{equation}	
	\begin{aligned}	
		A_{a}&=	\frac{g^{\rho ^0\rightarrow \pi ^+\pi ^- }}{s-m_{\rho^0}^{2}+im_{\rho^0} \varGamma _{\rho^0}}{\cal A}(\rho ^0{\pi}).
	\end{aligned}
\end{equation}

Based on the contributions depicted in diagram (e) and diagram (h)  within the framework of PQCD approach, we have derived the decay amplitudes $A_{e}$ and $A_{h}$ that encompass mixed contributions, expressed as follows:
	\begin{equation}
	\begin{aligned}
		A_{e}&=\frac{g^{\rho ^0\rightarrow \pi ^+\pi ^- }}{(s-m_{\rho^0}^{2}+im_{\rho^0} \varGamma _{\rho^0})({s-m_{\phi}^{2}+im_{\phi} \varGamma_{\phi}})}\overset{\sim}{\Pi}_{\mathrm{\rho^0\phi}}{\cal A}(\phi{\pi}),
	\end{aligned}
\end{equation}	
	\begin{equation}
		\begin{aligned}
		A_{h}&=\frac{g^{\rho^0\rightarrow \pi ^+\pi ^- }}{(s-m_{\rho^0}^{2}+im_{\rho^0} \varGamma _{\rho^0})({s-m_{\omega}^{2}+im_{\omega} \varGamma_{\omega}})}\overset{\sim}{\Pi}_{\mathrm{\rho^0\omega}}{\cal A}(\omega{\pi}),
		\end{aligned}
	\end{equation}
where  ${\cal A}(\phi{\pi})$ and ${\cal A}(\omega{\pi})$ respectively represent the decay amplitudes of the $B\rightarrow\phi\pi$ and $B\rightarrow\omega\pi$. 
	We define the differential parameter of CP asymmetry as:
	\begin{equation}
		\label{cp31}
		A_{CP}=\frac{\left| A \right|^2-\left| \overline{A} \right|^2}{\left| A \right|^2+\left| \overline{A} \right|^2}.
	\end{equation}

	\subsection{\label{subsec:form}  Regional CP asymmetry}
Recently, the LHCb experiment has made a significant discovery by directly measuring CP asymmetry in charmless decays of B meson involving three-body final states, providing evidence for substantial values within a specific region of phase space.		
	The decay process ${B}_{u}^{-}\rightarrow\pi ^+\pi ^-\pi^{-}$ exhibits a significant CP asymmetry within the invariant mass region ($m^2_{\pi^+\pi^- \rm ~low}<0.4~{\rm GeV}^2 $) \cite{LHCb:2014mir}. In this study, we compute the $A_{CP}$ integrals to facilitate comparison with experimental observations.

	 The amplitude of the ${B}_{u}^{-}\rightarrow \rho^0 \pi^{-}$ decay can be expressed as $ M_{{B}_{u}^{-}\rightarrow \rho^0 \pi^{-}}^{\lambda}=\alpha P_{{B}_{u}^{-}}\cdot \epsilon ^*\left( \lambda \right),$ where $\alpha$ is an input parameter. $P_{{B}_{u}^{-}}$ represents the momentum of ${B}_{u}^{-}$, $\epsilon $ denotes the polarization vector of $\rho^0$ and $\lambda$ signifies the polarization direction of $\epsilon $.
     The amplitude for $\rho^0 \rightarrow \pi ^+\pi ^-$ can be written as follows:
	$M_{\rho ^0\rightarrow \pi ^+\pi ^-}^{\lambda}={g^{\rho ^0\rightarrow \pi ^+\pi ^- }}\epsilon \left( \lambda \right) \cdot \left( p_\pi ^+-p_\pi ^- \right)$, where ${g^{\rho ^0\rightarrow \pi ^+\pi ^- }}$ is the effective coupling constant
	and $\epsilon$ refers to the plolarization vector of $\rho^0$. Hence, the amplitude of ${B}_{u}^{-}\rightarrow \rho^0(\rho^{0}\rightarrow\pi ^+\pi ^-)\pi^-$ is:
	\begin{equation}
		\begin{aligned}
			A&=\frac{{g^{\rho ^0\rightarrow \pi ^+\pi ^- }}\alpha}{(s-m_{\rho^0}^{2}+im_{\rho^0} \varGamma _{\rho^0})}P_{{B}_{u}^{-}}^{\mu}\sum_{\lambda =\pm 1,0}{\epsilon _{\mu}^{*}\left( \lambda \right) \epsilon _r\left( \lambda \right) \cdot \left( p_\pi ^+-p_\pi ^- \right) ^r}
			\\ &
			=-\frac{{g^{\rho ^0\rightarrow \pi ^+\pi ^- }}\alpha}{(s-m_{\rho^0}^{2}+im_{\rho^0} \varGamma _{\rho^0})}P_{{B}_{u}^{-}}^{\mu}\left[ g_{\mu r}-\frac{\left(  p_\pi ^++p_\pi ^- \right) _{\mu}\left(  p_\pi ^++p_\pi ^- \right) _r}{m_{\rho^0}^{2}} \right] \left(  p_\pi ^+-p_\pi ^-  \right) ^r		
			\\ &
			=\frac{{g^{\rho ^0\rightarrow \pi ^+\pi ^- }}}{(s-m_{\rho^0}^{2}+im_{\rho^0} \varGamma _{\rho^0})} \cdot \frac{M_{{B}_{u}^{-}\rightarrow \rho^0 \pi^{-}}^{\lambda}}{p_{{B}_{u}^{-}} \cdot \epsilon^{*}} \cdot\left(\Sigma-s^{\prime}\right). 
		\end{aligned}
	\end{equation}
	
	In the three-body decay process, we utilize the principles of momentum and energy conservation to transform the equation as follows: $P_{{B}_{u}^{-}}= p_\pi ^++p_\pi ^- +p_\pi ^-$ and $m_{ij}^{2}=p_{ij}^{2}$, aiming to facilitate computational procedures. Here, $\sqrt{s}$ and $\sqrt{s^{\prime}}$ represent the low and high invariant mass of the pair of $\pi^{+}\pi^{-}$ mesons respectively.
	By fixing the variable $s$ and considering the dependency between its values, we can determine an appropriate value $s^{'}$ that satisfies the formula $\Sigma=\frac{1}{2}\left(s_{\max }^{\prime}+s_{\min }^{\prime}\right)$, where $s_{\max}^{'}$ and $s_{\min}^{'}$ represent the maximum and minimum values respectively \cite{Qi:2018lxy}.	
	According to the kinetic principles of the three-body decay process, we can derive the regional CP asymmetry in the ${B}_{u}^{-}\rightarrow\pi ^+\pi ^-\pi^{-}$ decay process within a certain invariant mass ranges:
	
	\begin{eqnarray}
		A_{CP}^{\varOmega}=\frac{\int_{s_1}^{s_2}{ds\int_{s_{1}^{'}}^{s_{2}^{'}}{ds^{'}}}\left( \Sigma -s^{'} \right) ^2\left( \left|  {A} \right|^2-\left| \overline{ {A}} \right|^2 \right)}{\int_{s_1}^{s_2}{ds\int_{s_{1}^{'}}^{s_{2}^{'}}{ds^{'}}}\left( \Sigma -s^{'} \right) ^2\left( \left|  {A} \right|^2+\left| \overline{ {A}} \right|^2 \right)}.
	\end{eqnarray}

	The numerator and denominator of $A_{CP}$ can be integrated over the range $\Omega \left(s_{1}<s<s_{2}, s_{1}^{\prime}<s^{\prime}< s_{2}^{\prime}\right)$. The integration interval for high-invariant mass of $\pi ^+\pi ^-$ meson pair is $ s_{1}^{\prime}<s^{\prime}< s_{2}^{\prime}$, where $\int_{s_{1}^{'}}^{s_{2}^{'}}{ds'}\left( \Sigma -s' \right) ^2$ represents a factor dependent on $s'$.
	Through kinematic analysis, the correlation between $ \Sigma$ and $s^{\prime}$ can be readily determined. Assuming a finite range, we can consider $ \Sigma$ to be constant. Consequently, the term $\int_{s_{1}^{'}}^{s_{2}^{'}}{ds^{'}}\left( \Sigma  -s^{'} \right) ^2$ becomes negligible in the calculation, rendering $A_{CP}^{\varOmega}$ independent of the high invariant mass of positive and negative particles. In the $B\rightarrow\pi ^+\pi ^-K$ decay process, the similar method can also be used to calculate regional CP asymmetry.

	\section{\label{cal}Calculation process and Analysis of CP asymmetry}

	 In the calculation of CP asymmetry, we consider the contribution of the mixed resonance mechanism to the three-body decay amplitude. For B meson decays, the PQCD method is employed to separate the decay process into its hard components and non-perturbative parts. The hard components are isolated and analyzed using perturbative theory, while the non-perturbative parts are incorporated into universal meson wave functions.	
The decay of B mesons involves intricate dynamical phenomena, wherein a light quark exhibits significant kinetic energy while the spectator quark remains relatively stationary. Subsequently, the spectator quark undergoes an exchange of a high-energy gluon, thereby acquiring additional kinetic energy and accelerating to combine with a light quark, resulting in the production of a rapidly moving final meson.	
	The final meson further decays into the pair of $\pi^{+}\pi^{-}$ mesons, and the transverse momentum $(k_T)$ is retained in the PQCD method. To handle endpoint divergences, the Sudakov factor is introduced to suppress long-range interactions in the small transverse momentum range, thus ensuring that the entire process can be effectively perturbatively calculated \cite{Li:2004ep,Chen:2002th}. 
	The parameters used in the calculation are derived from Table \ref{tab:constant} .
	
	\begin{table}[h]
		\caption{Other parameters are from \ \cite{Yang:2022ebu,ParticleDataGroup:2024cfk}}
		\label{tab:constant}
		\centering
		\begin{tabular*}{18cm}{@{\extracolsep{\fill}}lllll}
			\hline\hline\\
			\text{Mass(\text{GeV})}
			& $m_{B_u}=5.27934$    & $m_{B_d}=5.27965$  &$m_{\pi^{\pm}}=0.140$ &$m_{\pi^{0}}=0.135$\\[1ex]
			\text{Wolfenstein parameters}
			& $\lambda =0.22650$  & $A=0.790$  &$\bar{\rho}=0.141$ & $\bar{\eta}=0.357$ \\[1ex]
			\text{Decay constants (GeV)}
			& $f_{B_s}=0.23$ & $f_{B}=0.21$  &$f_{\rho(770)}=0.216$ &$f_{\rho(1020)}^T=0.184$ \\[1ex]
			\text{Decay width (\text{GeV})}
			& $\Gamma_{\rho}$ = $0.15 $ & $\Gamma_{\omega}$ = $8.49 \times 10^{-3}$ &$ \Gamma_{\phi}$ = $4.23 \times 10^{-3} $\\[1ex]
			\hline\hline
		\end{tabular*}
	\end{table}
	
	\subsection{ CP asymmetry analysis of the ${B}_u^{-}\rightarrow (\rho^{0 },\omega ,\phi\rightarrow \pi^{+}\pi^{-})\pi^{-}$  decay process }
	
	The Dalitz diagram analysis of the decay amplitude $ {B}^{ \pm} \rightarrow \pi^{+}\pi^{-}\pi^{\pm}$ in two-dimensional phase space is performed, as indicated by relevant research literature \cite{LHCb:2013lcl}. Phenomenological investigations have primarily focused on exploring regional CP asymmetries in the decay process $ {B}^{ \pm} \rightarrow \pi^{+}\pi^{-}\pi^{\pm}$, with certain studies emphasizing the significance of $\rho^{0 }-\omega $ mixing effects between $\rho^{0 }$ and $\omega $, highlighting potential interference between the resonance of $\rho^{0 }$ and the broad S-wave contribution \cite{LHCb:2014mir,LHCb:2013lcl}.             	Q
	The resonance contributions of $\rho-\omega$, $\omega-\phi$ and $\rho-\phi$ can give rise to a novel strong phase. 
	To facilitate clearer observation, we have selected the energy range of 0.7 GeV to 1.1 GeV, which corresponds to the primary region where significant effects of decay processes involving $\rho^{0}$, $\omega$ and $\phi$ resonances are exhibited.

	We can obtain Dalitz diagram for the B meson decay process from experiments. By analyzing the Dalitz diagram, we can determine the existence of resonance states, study the energy momentum relationships in the decay process, and further explore the impact of resonance and non-resonance contributions on the B meson three-body decay process. However, due to the fact that the B meson phase space allows for multiple types of resonances to exist, there may be a large number of intermediate states. Investigating these resonance structures is a complex task and it should be noted that it is not possible to distinguish between the contributions made by $\rho^{0}$ meson and $\omega$ meson in experiments.
	
Using the PQCD method and taking into account the CKM matrix elements $V_{ub} V_{ud}^{*}$ and $V_{tb} V_{td}^{*}$ in  $B_{u}^{-}\rightarrow (\rho^{0}, \rightarrow \pi^{+}\pi^{-})\pi^{-}$ three-body decay process, we can give the amplitude expression for CP asymmetry. 
Fig. 1 encompasses the direct decay modes of vector meson and the mixed resonance decay modes of two vector mesons. The three-body decay amplitude of the B meson in diagram (a) of Fig. 1 is expressed as follows:  
	
	\begin{equation}
	\begin{aligned}
	{ A}(a)=	{ A}(B_{u}^{-}{\to}(\rho ^0\rightarrow \pi ^+\pi ^- ){\pi}^{-}) 
	&=\sum_{\lambda =0,\pm 1} \frac{G_{F}P_{B_{u}^{-}}\cdot \epsilon^*\left( \lambda \right)\ g^{\rho ^0\rightarrow \pi ^+\pi ^- }\epsilon \left( \lambda \right) \cdot \left( p_{\pi ^+}-p_{\pi ^-} \right)}{{2}({s-m_{\rho ^0}^{2}+im_{\rho ^0} \varGamma _{\rho ^0}})} 
		\\& 
		\times \bigg\{V_{ub}\,V_{ud}^{\ast}\, \big\{
		a_{1}\, \big[
		{\cal A}_{ab}^{LL}({\pi},{\rho})
		+ {\cal A}_{ef}^{LL}({\pi},{\rho})
		- {\cal A}_{ef}^{LL}({\rho},{\pi}) \big]
		+ a_{2}\, {\cal A}_{ab}^{LL}({\rho},{\pi})
		\\ & 
		+ C_{2}\, \big[
		{\cal A}_{cd}^{LL}({\pi},{\rho})
		+ {\cal A}_{gh}^{LL}({\pi},{\rho})
		- {\cal A}_{gh}^{LL}({\rho},{\pi}) \big]
		+ C_{1}\, {\cal A}_{cd}^{LL}({\rho},{\pi}) \big\}
		- V_{tb}\,V_{td}^{\ast}\, \big\{
		(a_{4} 
		\\ &
		+a_{10}) \, \big[
		{\cal A}_{ab}^{LL}({\pi},{\rho})
		+ {\cal A}_{ef}^{LL}({\pi},{\rho})
		- {\cal A}_{ef}^{LL}({\rho},{\pi}) \big]
		+ (a_{6}+a_{8})\, \big[
		{\cal A}_{ab}^{SP}({\pi},{\rho})
		\\ & 
		+ {\cal A}_{ef}^{SP}({\pi},{\rho})
		- {\cal A}_{ef}^{SP}({\rho},{\pi}) \big]
		- (a_{4} -\frac{3}{2}\,a_{7}
		- \frac{3}{2}\,a_{9} -\frac{1}{2}\,a_{10} ) \,
		{\cal A}_{ab}^{LL}({\rho},{\pi})
		+ (C_{3}
		\\ &
		+C_{9}) \, \big[
		{\cal A}_{cd}^{LL}({\pi},{\rho})
		+ {\cal A}_{gh}^{LL}({\pi},{\rho})
		- {\cal A}_{gh}^{LL}({\rho},{\pi}) \big]
		+ (C_{5}
		+C_{7})\, \big[
		{\cal A}_{cd}^{SP}({\pi},{\rho})
		\\ &
		+ {\cal A}_{gh}^{SP}({\pi},{\rho})
		- {\cal A}_{gh}^{SP}({\rho},{\pi}) \big]
		- (C_{3} - \frac{3}{2}\,C_{10} -\frac{1}{2}\,C_{9} ) \,
		{\cal A}_{cd}^{LL}({\rho},{\pi})
		+ \frac{3}{2}\,C_{8}\,
		{\cal A}_{cd}^{LR}({\rho},{\pi})
		\\ &  
		- (C_{5}-\frac{1}{2}\,C_{7})\,
		{\cal A}_{cd}^{SP}({\rho},{\pi}) \big\}
		\label{amp-pim-rhoz}\bigg\}.
	\end{aligned}
\end{equation}
	The momentum parameter $P_{i}$ ($P_{i}=P_{{B}_{u}^{-}}$, $p_{\pi ^+}$, $p_{\pi ^-}$) is defined, where $\epsilon$ represents the polarization vector of the vector meson. The Fermi coupling constant is denoted as $G_F$. $C_{i}$ is Wilson coefficient, $a_{i}$ is related to Wilson coefficient $C_{i}$ \cite{Li:2006jv}.  Three flow structures are labeled as $LL, LR,$ and $SP$. ${\cal A}_{ab}$ refers to the contribution of factorizable emission diagrams, while ${\cal A}_{cd }$ represents the contribution of nonfactorizable emission diagrams. Similarly, ${\cal A}_{ef}$ (${\cal A}_{gh}$) denotes  the contribution of factorizable (nonfactorizable) annihilation diagrams. 	
	 $V_{ub} V_{ud}^{*}$ and $V_{tb} V_{td}^{*}$ can be measured experimentally and theoretically represented by the Wolfstein parameters $A$, $\rho$, $\lambda$ and $\eta$: $V_{ub} V_{ud}^{*}=A \lambda^{3}(\rho-i\eta)(1-\frac{\lambda^2}{2})$ and $V_{tb} V_{td}^{*}=A \lambda^{3}(1-\rho+i\eta)$.

Diagram (d) and diagram (g)  of Fig. 1 respectively represent the direct decay process of the $\phi$ and $\omega$ mesons, producing the pair of $\pi^+\pi^-$ mesons. 
Due to isospin symmetry breaking, the contribution of $\omega$ and $\phi$ to this decay can be neglected. 
The specific expression of the three-body decay amplitude in the vector meson resonance, as shown in diagram (e):
	\begin{equation}
	\begin{aligned}
	{ A}(e)=	{ A}(B_{u}^{-}{\to}(\phi-\rho ^0\rightarrow \pi ^+\pi ^- ){\pi}^{-})
		&	
		=\sum_{\lambda =0,\pm 1} \frac{-G_{F}P_{{B}_{u}^{-}}\cdot \epsilon^*\left( \lambda \right)\ g^{\rho^0\rightarrow \pi ^+\pi ^- }\epsilon \left( \lambda \right) \cdot \left( p_{\pi ^+}-p_{\pi ^-} \right)}{{2}(s-m_{\rho^0}^{2}+im_{\rho^0} \varGamma _{\rho^0})({s-m_{\phi}^{2}+im_{\phi} \varGamma_{\phi}})}\overset{\sim}{\Pi}_{\mathrm{\rho^0\phi}}
		\\	&
		\times \bigg\{ V_{tb}\,V_{td}^{\ast}\, \big\{
		(a_{3}+a_{5}-\frac{1}{2}\,a_{7}-\frac{1}{2}\,a_{9}) \,
		{\cal A}_{ab}^{LL}({\phi},{\pi}) 
		+ (C_{4}
		-\frac{1}{2}\,C_{10})\,
		{\cal A}_{cd}^{LL}({\phi},{\pi})
		\\ &
		+ (C_{6}-\frac{1}{2}\,C_{8})\,
		{\cal A}_{cd}^{LR}({\phi},{\pi}) \bigg\}.
	\end{aligned}
\end{equation}

The $\rho^0$ meson decays into the pair of $\pi^+\pi^-$ mesons through $\omega$ meson resonance, the amplitude of the three-body decay shown in diagram (h) of Fig. 1 is as follows:	
\begin{equation}
	\begin{aligned}
	{ A}(h)=	{ A}(B_{u}^{-}{\to}(\omega-\rho ^0\rightarrow \pi ^+\pi ^- ){\pi}^{-})
		&
		=\sum_{\lambda =0,\pm 1} \frac{G_{F}P_{{B}_{u}^{-}}\cdot \epsilon^*\left( \lambda \right)\ g^{\rho^0\rightarrow \pi ^+\pi ^- }\epsilon \left( \lambda \right) \cdot \left( p_{\pi ^+}-p_{\pi ^-} \right)}{{2}(s-m_{\rho^0}^{2}+im_{\rho^0} \varGamma _{\rho^0})({s-m_{\omega}^{2}+im_{\omega} \varGamma_{\omega}})}\overset{\sim}{\Pi}_{\mathrm{\rho^0\omega}}
		\\	&
		\times \bigg\{ V_{ub}\,V_{ud}^{\ast}\, \big\{
		a_{1}\, \big[
		{\cal A}_{ab}^{LL}({\pi},{\omega})
		+ {\cal A}_{ef}^{LL}({\pi},{\omega})
		+ {\cal A}_{ef}^{LL}({\omega},{\pi}) \big]
		+ a_{2}\, {\cal A}_{ab}^{LL}({\omega},{\pi})
		\\ &  
		+ C_{2}\, \big[
		{\cal A}_{cd}^{LL}({\pi},{\omega})
		+ {\cal A}_{gh}^{LL}({\pi},{\omega})
		+ {\cal A}_{gh}^{LL}({\omega},{\pi}) \big]
		+ C_{1}\, {\cal A}_{cd}^{LL}({\omega},{\pi}) \big\}
		\\ &- V_{tb}\,V_{td}^{\ast}\, \big\{
		(a_{4}+a_{10}) \, \big[
		{\cal A}_{ab}^{LL}({\pi},{\omega})
		+ {\cal A}_{ef}^{LL}({\pi},{\omega})
		+ {\cal A}_{ef}^{LL}({\omega},{\pi}) \big]
		\\  & 
		+ (a_{6}+a_{8})\, \big[
		{\cal A}_{ab}^{SP}({\pi},{\omega})
		+ {\cal A}_{ef}^{SP}({\pi},{\omega})
		+ {\cal A}_{ef}^{SP}({\omega},{\pi}) \big]
		\\ & 
		+ (2\,a_{3}+a_{4} +2\,a_{5}+ \frac{1}{2}\,a_{7}
		+ \frac{1}{2}\,a_{9} -\frac{1}{2}\,a_{10} ) \,
		{\cal A}_{ab}^{LL}({\omega},{\pi})
		\\  & \
		+ (C_{3}+C_{9}) \, \big[
		{\cal A}_{cd}^{LL}({\pi},{\omega})
		+ {\cal A}_{gh}^{LL}({\pi},{\omega})
		+{\cal A}_{gh}^{LL}({\omega},{\pi}) \big]
		\\ &  
		+(C_{5}+C_{7})\, \big[
		{\cal A}_{cd}^{SP}({\pi},{\omega})
		+ {\cal A}_{gh}^{SP}({\pi},{\omega})
		+ {\cal A}_{gh}^{SP}({\omega},{\pi}) \big]
		\\ &  
		+ (C_{3} +2\,C_{4} - \frac{1}{2}\,C_{9} + \frac{1}{2}\,C_{10} ) \,
		{\cal A}_{cd}^{LL}({\omega},{\pi})
		\\ & 
		+ (2\,C_{6}+\frac{1}{2}\,C_{8} )\,
		{\cal A}_{cd}^{LR}({\omega},{\pi})
		+ (C_{5}-\frac{1}{2}\,C_{7})\,
		{\cal A}_{cd}^{SP}({\omega},{\pi}) \big\}
		\label{amp-pim-omega}\bigg\}.
	\end{aligned}
\end{equation}

The contributions of the mixed resonance modes $\rho^{0}-\phi$, $\omega-\phi$, $\rho^{0}-\omega$, and $\phi-\omega$ to the pair of $\pi^+\pi^-$ mesons are comparatively smaller than those of the direct decay processes, namely, $\phi \rightarrow \pi\pi$ and $\omega \rightarrow \pi\pi$. Hence, we do not consider the contributions from diagram (b), diagram (c), diagram (f), and diagram (i) in Fig. 1.

	\begin{figure}[!htbp]
		\centering
		\begin{minipage}[h]{0.45\textwidth}
			\centering
			\includegraphics[height=4cm,width=6.5cm]{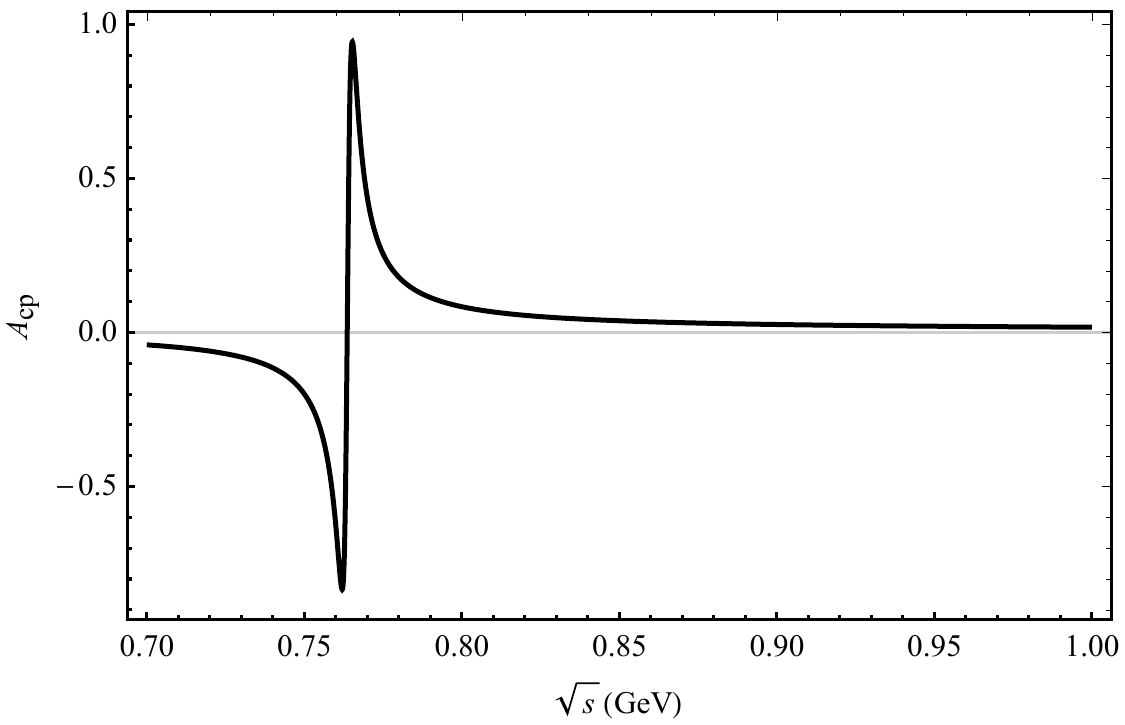}
			\caption{Plot of  $A_{CP}$ as a function of $\sqrt{s}$ corresponding to central parameter values of CKM matrix elements
				for the decay channel of $ {B}^{-} \rightarrow \pi^{+}\pi^{-}\pi^{-}$.}
			\label{fig2}
		\end{minipage}
		\quad
		\begin{minipage}[h]{0.45\textwidth}
			\centering
			\includegraphics[height=4cm,width=6.5cm]{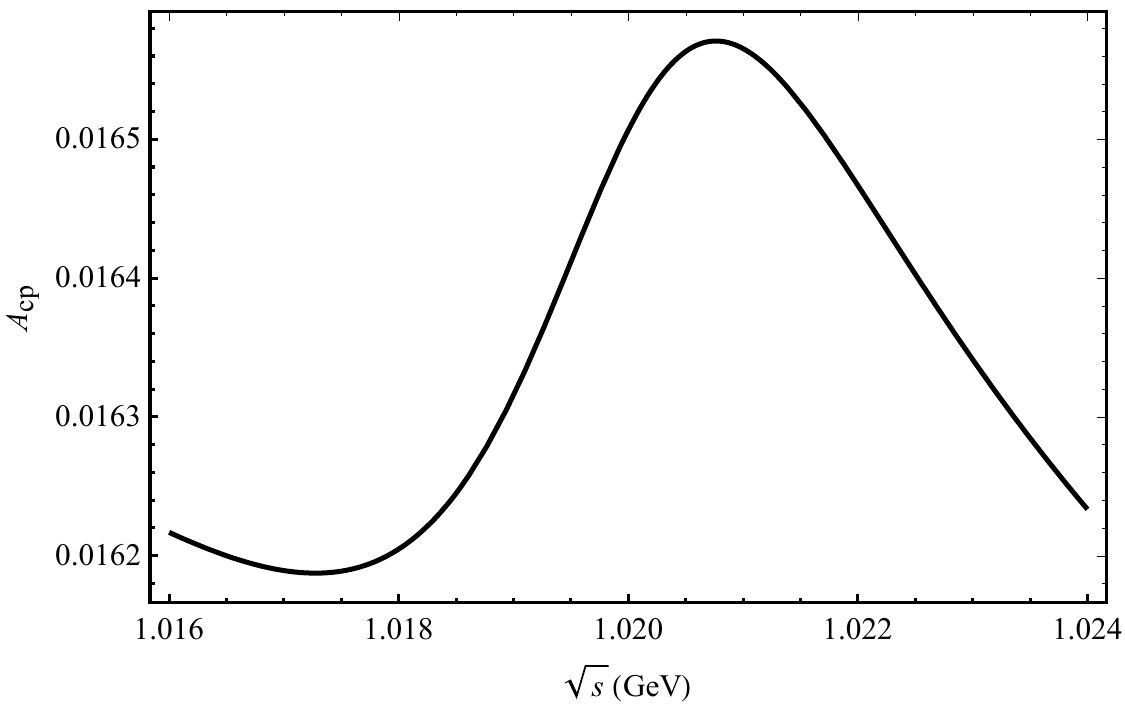}
			\caption{CP asymmetry plot near the invariant mass of $\phi$ during the decay channel of  $ {B}^{-} \rightarrow \pi^{+}\pi^{-}\pi^{-}$
				from the central parameter values of CKM matrix elements.}		
			\label{fig3}
		\end{minipage}
	\end{figure}
	We present a diagram illustrating the relationship between CP asymmetry and the invariant mass $\sqrt{s}$ in the quasi-two-body decay $B_{u}^{-}\rightarrow (\rho^{0},\omega, \phi\rightarrow \pi ^+\pi ^-)\pi^-$ within the interference range.
	The results for the $B_{u}^{-}\rightarrow \pi^{+}\pi^{-}\pi^{-}$ decay process are presented in Fig. 2 and Fig. 3. As depicted in Fig. 2 and Fig. 3, it is evident that the CP asymmetry of the $B_{u}^{-}\rightarrow \pi^{+}\pi^{-}\pi^{-}$ channel undergoes a change when the invariant masses of the $\pi^+\pi^{-}$ pair encompass the resonance range of $\omega$ and $\phi$, with a maximum CP asymmetry of $94\%$. The decay process $B_{u}^{-}\rightarrow \pi^{+}\pi^{-}\pi^{-}$ exhibits a significant variation in CP asymmetry when the invariant masses of the $\pi^{+}\pi^{-}$ pair approaches 0.76 GeV, reaching a peak value of $94\%$. This behavior can be attributed to the effect arising from the mixing mechanism between $\rho^{0 }$ and $\omega$. Consequently, interference effects are expected within the range of 0.7 GeV-0.8 GeV as indicated by Fig. 2, along with small peaks observed in the invariant mass range corresponding to $\phi$ according to Fig. 3 which CP asymmetry is measured at $1.6\%$.

	\subsection{CP asymmetry analysis of the ${B}_u^{-}\rightarrow (\rho^{0 },\omega, \phi\rightarrow \pi^{+}\pi^{-}) K^{-}$  decay process }
	
	 Experimental observations have revealed that specific regional CP asymmetries in phase space exhibit a greater magnitude compared to the overall CP asymmetry in phase space \cite{Cheng:2016shb}.
The CP asymmetry in the $B^{-}{\to} \pi ^+\pi ^-  K^{-}$ decay process has been measured to be $2.5\pm0.4\pm0.4\pm0.7 \%$ in the full phase space and  $67.8\pm7.8\pm3.2\pm0.7 \%$ within a specified low invariant mass region  \cite{LHCb:2013ptu, LHCb:2014mir}.

	  Considering the resonance effect of vector mesons, namely $\rho^{0}$, $\omega$ and $\phi$, we present the amplitude of the decay process ${B}_u^{-}\rightarrow (\rho^{0}, \omega, \phi \rightarrow \pi^{+}\pi^{-})K^{-}$ following a similar mechanism. The three-body direct decay amplitude in diagram (a) of Fig. 1 is as follows: 
	  \begin{equation}
	  	\begin{aligned}
	  		{ A}(a)=	{A}\left(B_{u}^{-}{\to}\left(\rho ^0\rightarrow \pi ^+\pi ^-\right )K^{-}\right)
	  		&	=\sum_{\lambda =0,\pm 1} \frac{G_{F}P_{B_{u}^{-}}\cdot \epsilon^*\left( \lambda \right)\ g^{\rho ^0\rightarrow \pi ^+\pi ^- }\epsilon \left( \lambda \right) \cdot \left( p_{\pi ^+}-p_{\pi ^-} \right)}{{2}({{s-m_{\rho ^0}^{2}+im_{\rho ^0} \varGamma _{\rho ^0}}})} 
	  		\\& 
	  		\times \bigg\{V_{ub}\,V_{us}^{\ast}\big\{ a_{1}\, \big[	{\cal A}_{ab}^{LL}({K},{\rho})	+ {\cal A}_{ef}^{LL}({K},{\rho}) \big]	+ a_{2}\,{\cal A}_{ab}^{LL}({\rho},{K})
	  		+C_{2} \big[{\cal A}_{cd}^{LL}({K},{\rho})
	  		\\ &
	  		+ {\cal A}_{gh}^{LL}({K},{\rho}) \big]	+ C_{1}\,{\cal A}_{cd}^{LL}({\rho},{K}) \big\}	-	V_{tb}\,V_{ts}^{\ast}\, \big\{( a_{4}
	  		+a_{10} )\, \big[{\cal A}_{ab}^{LL}({K},{\rho})	+ {\cal A}_{ef}^{LL}({K},{\rho}) \big]
	  		\\ &
	  		+  ( a_{6}+a_{8} )\, \big[	{\cal A}_{ab}^{SP}({K},{\rho})
	  		+  	{\cal A}_{ef}^{SP}({K},{\rho}) \big]	+ \frac{3}{2}\, (a_{7}+a_{9})\,	{\cal A}_{ab}^{LL}({\rho},{K})	+	( C_{3}
	  		\\ &
	  		+C_{9} )\, \big[	{\cal A}_{cd}^{LL}({K},{\rho})+
	  		{\cal A}_{gh}^{LL}({K},{\rho}) \big]	+  ( C_{5}+C_{7} )\, \big[	{\cal A}_{cd}^{SP}({K},{\rho})	+ {\cal A}_{gh}^{SP}({K},{\rho}) \big]
	  		\\ &
	  		+ \frac{3}{2}\, C_{8}\,{\cal A}_{cd}^{LR}({\rho},{K})	+ \frac{3}{2}\, C_{10}\,{\cal A}_{cd}^{LL}({\rho},{K}) \big\}
	  		\label{km-rhoz-amp}\bigg\}.
	  	\end{aligned}	
	  \end{equation}
	  	  
	The $\rho^0$ meson decays into the pair of $\pi^+\pi^-$ mesons through $\phi$ meson resonance, the amplitude of the three-body decay shown diagram (e) of Fig. 1 is as follows:	
	\begin{equation}
		\begin{aligned}
			{ A}(e)=	{ A}(B_{u}^{-}{\to}(\phi-\rho^0\rightarrow \pi ^+\pi ^- ) K^{-})
			&		
			=\sum_{\lambda =0,\pm 1} \frac{G_{F}P_{{B}_{u}^{-}}\cdot \epsilon^*\left( \lambda \right)\ g^{\rho^0\rightarrow \pi ^+\pi ^- }\epsilon \left( \lambda \right) \cdot \left( p_{\pi ^+}-p_{\pi ^-} \right)}{\sqrt {2}(s-m_{\rho^0}^{2}+im_{\rho^0} \varGamma _{\rho^0})({s-m_{\phi}^{2}+im_{\phi} \varGamma_{\phi}})}\overset{\sim}{\Pi}_{\mathrm{\rho^0\phi}} \\&
			\times \bigg\{ V_{ub}\,V_{us}^{\ast}\, \big\{
			a_{1}\, {\cal A}_{ef}^{LL}({\phi},{K})
			+C_{2}\, {\cal A}_{gh}^{LL}({\phi},{K}) \big\}
			- V_{tb}\,V_{ts}^{\ast}\, \big\{
			( a_{3}+a_{4}+a_{5}-\frac{1}{2}\,a_{7}\\&
			-\frac{1}{2}\,a_{9}-\frac{1}{2}\,a_{10} )\,
			{\cal A}_{ab}^{LL}({\phi},{K})
			+  ( a_{4}+a_{10} )\,{\cal A}_{ef}^{LL}({\phi},{K})
			+  ( a_{6}+a_{8} )\,{\cal A}_{ef}^{SP}({\phi},{K})
          \\  &
			+ ( C_{3}+C_{4}-\frac{1}{2}\,C_{9}-\frac{1}{2}\,C_{10} )\,
			{\cal A}_{cd}^{LL}({\phi},{K})
			+ ( C_{6}-\frac{1}{2}\,C_{8} )\,
			{\cal A}_{cd}^{LR}({\phi},{K})
			+ ( C_{5} \\ &
			-\frac{1}{2}\,C_{7} )\,
			{\cal A}_{cd}^{SP}({\phi},{K})
			+  ( C_{3}+C_{9} )\,{\cal A}_{gh}^{LL}({\phi},{K})
			+  ( C_{5}+C_{7} )\,{\cal A}_{gh}^{SP}({\phi},{K})  \bigg\}
			\label{km-phi-amp}.
		\end{aligned}
	\end{equation}

	The $\rho^0$ meson decays into the pair of $\pi^+\pi^-$ mesons through $\omega$ meson resonance, the amplitude of the three-body decay shown diagram (h) of Fig. 1 is as follows:	
	\begin{equation}
	\begin{aligned}
		{ A}(h)=	{ A}(B_{u}^{-}{\to}(\omega-\rho^0\rightarrow \pi ^+\pi ^- )K^{-})
		&=\sum_{\lambda =0,\pm 1} \frac{G_{F}P_{{B}_{u}^{-}}\cdot \epsilon^*\left( \lambda \right)\ g^{\rho^0\rightarrow \pi ^+\pi ^- }\epsilon \left( \lambda \right) \cdot \left( p_{\pi ^+}-p_{\pi ^-} \right)}{{2}(s-m_{\rho^0}^{2}+im_{\rho^0} \varGamma _{\rho^0})({s-m_{\omega}^{2}+im_{\omega} \varGamma_{\omega}})}\overset{\sim}{\Pi}_{\mathrm{\rho\omega}} \\&
		\times \bigg\{ V_{ub}\,V_{us}^{\ast}\,\big\{ a_{1}\, \big[{\cal A}_{ab}^{LL}({K},{\omega})+ {\cal A}_{ef}^{LL}{K},{\omega}) \big]+ a_{2}\,{\cal A}_{ab}^{LL}({\omega},{K})\\ &
		+C_{2} \big[{\cal A}_{cd}^{LL}({K},{\omega})	+ {\cal A}_{gh}^{LL}({K},{\omega}) \big]	+ C_{1}\,{\cal A}_{cd}^{LL}({\omega},{K}) \big\}	-	V_{tb}\,V_{ts}^{\ast}\, \big\{( a_{4}
		\\ &
		+a_{10} )\, \big[{\cal A}_{ab}^{LL}({K},{\omega})+ {\cal A}_{ef}^{LL}({K},{\omega}) \big]
		+ ( a_{6}+a_{8} )\, \big[{\cal A}_{ab}^{SP}({K},{\omega}) \\ &+ {\cal A}_{ef}^{SP}({K},{\omega}) \big]+ (2\,a_{3}+2\,a_{5}+\frac{1}{2}\,a_{7}+\frac{1}{2}\,a_{9})\,{\cal A}_{ab}^{LL}({\omega},{K})+ ( C_{3}\\ & 
		+C_{9} )\, \big[{\cal A}_{cd}^{LL}({K},{\omega})+ {\cal A}_{gh}^{LL}({K},{\omega}) \big]+ ( C_{5}+C_{7} )\, \big[
		{\cal A}_{cd}^{SP}({K},{\omega})+ {\cal A}_{gh}^{SP}({K},{\omega}) \big]\\ & 
		+ (2\,C_{6}+\frac{1}{2}\, C_{8})\,{\cal A}_{cd}^{LR}({\omega},{K})+ (2\,C_{4}+\frac{1}{2}\, C_{10})\,{\cal A}_{cd}^{LL}({\omega},{K}) \big\}\label{km-w-amp}\bigg\},
	\end{aligned}
\end{equation}

	\begin{figure}[!htbp]
		\centering
		\begin{minipage}[h]{0.9\textwidth}
			\centering
			
			\includegraphics[height=6cm,width=9cm]{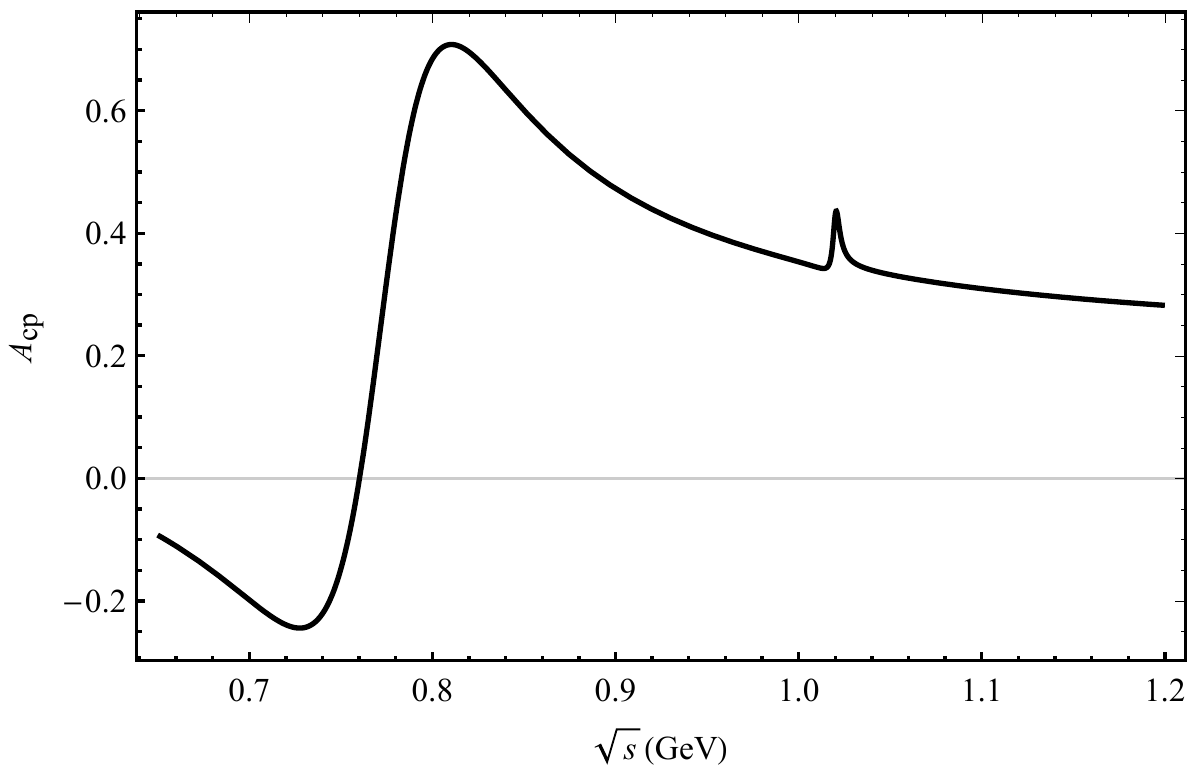}
			\caption{Plot of  $A_{CP}$ as a function of $\sqrt{s}$ corresponding to central parameter values of CKM matrix elements
				for the decay channel of ${B}_{u}^{-}\rightarrow \pi ^+\pi ^-K ^-$.}
			\label{fig1}
		\end{minipage}
		
	\end{figure}

	In Fig. 4, the observed behavior of the CP asymmetry in the $B_u^- \rightarrow \pi^+\pi^-K^-$ decay process provides valuable insights into the dynamics of this decay channel. As we approach the resonance range of $\omega$, a significant change in the CP asymmetry is evident, reaching a peak value of $70\%$. This suggests that there is a strong influence from the $\omega$ resonance on the decay process.	
	On the other hand, when we consider the resonance range of $\phi$, only slight variations in CP asymmetry are observed. This can be attributed to several factors. 	
	It is worth noting that the specific decay process $B_u^- \rightarrow (\phi\rightarrow\pi^+\pi^-) K^-$ experiences a further reduction in its overall amplitude due to interference from $\phi-\rho$ mixing resonance. This interference effect leads to a decrease in observable changes in CP asymmetry within the resonance range of $\phi$.

	\subsection{CP asymmetry analysis of the  $\bar{B}_d^{0}\rightarrow (\rho^{0}, \omega, \phi \rightarrow \pi ^+\pi ^-)\pi^{0}$  decay process  }

	Considering the vector meson $\rho^{0}$ - $\omega$ - $\phi$ resonance effect, the process of {$\bar{B}^{0}_d\rightarrow \pi ^+\pi ^-\pi^{0}$} can be further elaborated. The presence of these vector mesons in the decay channel introduces additional dynamics and interactions that contribute to the overall behavior of this process.
	They play a crucial role in understanding strong interactions. In particular, their resonant behavior is associated with a peak in the CP asymmetry for certain energy ranges.
	
	In the decay process of {$\bar{B}^{0}_d\rightarrow \pi ^+\pi ^-\pi^{0}$}, the vector meson resonance effect refers to how these particles can influence or modify the decay process.
	By considering this resonance effect, we gain insights into various aspects of {$\bar{B}^{0}_d\rightarrow \pi ^+\pi ^-\pi^{0}$}. For instance, it helps us understand how different intermediate states involving vector mesons contribute to the final state particles (pions) observed experimentally. It also provides information about possible interference patterns between different amplitudes contributing to this decay process.

	 The amplitudes of the decay process {$\bar{B}^{0}_d\rightarrow \pi ^+\pi ^-\pi^{0}$} arising from various intermediate vector mesons can be presented. The amplitude ${\cal A}({\bar{B}}^0_d \to (\rho^0 \rightarrow \pi^+\pi^-) \pi^0)$ corresponds to diagram (a) in Fig. 1, which can be expressed as:
		\begin{equation}
		\begin{aligned}
			{ A}(a)=	{ A}(\bar{B}^{0}_d
			{\to}(\rho ^0\rightarrow \pi ^+\pi ^- )\pi^{0})
			& 
			=\sum_{\lambda =0,\pm 1} \frac{G_{F}P_{\bar{B}^{0}}\cdot \epsilon^*\left( \lambda \right) g^{\rho ^0\rightarrow \pi ^+\pi ^- }\epsilon \left( \lambda \right) \cdot \left( p_{\pi ^+}-p_{\pi ^-} \right)}{2\sqrt{2}({{s-m_{\rho ^0}^{2}+im_{\rho ^0} \varGamma _{\rho ^0}}})}
			\\ &
			\times \bigg\{ V_{ub}\,V_{ud}^{\ast}\,
			\big\{ a_{2}\, \big[
			-{\cal A}_{ab}^{LL}({\pi},{\rho})
			-{\cal A}_{ab}^{LL}({\rho},{\pi})
			+{\cal A}_{ef}^{LL}({\pi},{\rho})
			+{\cal A}_{ef}^{LL}({\rho},{\pi}) \big]
			\\ & 
			+ C_{1}\, \big[
			-{\cal A}_{cd}^{LL}({\pi},{\rho})
			-{\cal A}_{cd}^{LL}({\rho},{\pi})
			+{\cal A}_{gh}^{LL}({\pi},{\rho})
			+{\cal A}_{gh}^{LL}({\rho},{\pi}) \big] \big\}
			\\ &-
			V_{tb}\,V_{td}^{\ast}\, \big\{
			( a_{4}-\frac{3}{2}\,a_{9}-\frac{1}{2}\,a_{10} )\, \big[
			{\cal A}_{ab}^{LL}({\pi},{\rho})
			+{\cal A}_{ab}^{LL}({\rho},{\pi}) \big]
			+ \frac{3}{2}\,a_{7}\, \big[
			{\cal A}_{ab}^{LL}({\pi},{\rho})
			\\ & 
			-{\cal A}_{ab}^{LL}({\rho},{\pi}) \big]
			+( C_{3}-\frac{1}{2}\,C_{9}-\frac{3}{2}\,C_{10} )\, \big[
			{\cal A}_{cd}^{LL}({\pi},{\rho})
			+{\cal A}_{cd}^{LL}({\rho},{\pi}) \big]
			\\ & 
			- \frac{3}{2}\,C_{8}\, \big[
			{\cal A}_{cd}^{LR}({\pi},{\rho})
			+{\cal A}_{cd}^{LR}({\rho},{\pi}) \big]		
			+ (a_{6}-\frac{1}{2}\,a_{8})\, \big[
			{\cal A}_{ab}^{SP}({\pi},{\rho})
			\\ & 
			+ {\cal A}_{ef}^{SP}({\pi},{\rho})	
			+ {\cal A}_{ef}^{SP}({\rho},{\pi}) \big]
			+ (C_{5}
			-\frac{1}{2}\,C_{7})\, \big[
			{\cal A}_{cd}^{SP}({\pi},{\rho})				
			+ {\cal A}_{cd}^{SP}({\rho},{\pi})	+ {\cal A}_{gh}^{SP}({\pi},{\rho})
			\\ & 
			+ {\cal A}_{gh}^{SP}({\rho},{\pi})
			+(2\,a_{3}+a_{4}-2\,a_{5}-\frac{1}{2}\,a_{7}
			+\frac{1}{2}\,a_{9}
			-\frac{1}{2}\,a_{10} )\, \big[
			{\cal A}_{ef}^{LL}({\pi},{\rho})
			\\ & 
			+{\cal A}_{ef}^{LL}({\rho},{\pi}) \big]		
			+(C_{3}+2\,C_{4}-\frac{1}{2}\,C_{9}+\frac{1}{2}\,C_{10} )\, \big[
			{\cal A}_{gh}^{LL}({\pi},{\rho})
			\\ &
			+{\cal A}_{gh}^{LL}({\rho},{\pi}) \big]
			+ ( 2\,C_{6}+\frac{1}{2}\,C_{8} )\, \big[
			{\cal A}_{gh}^{LR}({\pi},{\rho})
			+{\cal A}_{gh}^{LR}({\rho},{\pi}) \big] \big\}
			\label{piz-rhoz-amp}\bigg\},
		\end{aligned}
	\end{equation}

		The vector meson effective mixing resonance modes for the decay process ${\bar{B}}^0_d \to (\rho^0 \rightarrow \pi^+\pi^-) \pi^0$ are  $\phi-\rho^0$ and $\omega-\rho^0$, where the $\rho^0$ meson is produced through the resonant decay of the $\phi$ or $\omega$ meson to produce the pair of $\pi^+\pi^-$ mesons, corresponding to diagram (e) and diagram (h) of Fig. 1. The three-body decay amplitude forms of the mixing resonance modes are shown as:	
	
	\begin{equation}
		\begin{aligned}
			{ A}(e)=	{ A}(\bar{B}^{0}_d{\to}(\phi-\rho^0\rightarrow \pi ^+\pi ^- )\pi^{0})
			 &=
			\sum_{\lambda =0,\pm 1} \frac{G_{F}P_{\bar{B}^{0}}\cdot \epsilon^*\left( \lambda \right)\ g^{\rho^0\rightarrow \pi ^+\pi ^- }\epsilon \left( \lambda \right) \cdot \left( p_{\pi ^+}-p_{\pi ^-} \right)}{2(s-m_{\rho^0}^{2}+im_{\rho^0} \varGamma _{\rho^0})({s-m_{\phi}^{2}+im_{\phi} \varGamma_{\phi}})} \overset{\sim}{\Pi}_{\mathrm{\rho^0\phi}}
			\\  &				
			\times \bigg\{ V_{tb}\,V_{td}^{\ast}\, \big\{
			( a_{3}+a_{5}-\frac{1}{2}\,a_{7}-\frac{1}{2}\,a_{9} )\,
			{\cal A}_{ab}^{LL}({\phi},{\pi})
			+ ( C_{4}-\frac{1}{2}\,C_{10} )\,
			{\cal A}_{cd}^{LL}({\phi},{\pi})
			\\ & 
			+ ( C_{6}-\frac{1}{2}\,C_{8})\,
			{\cal A}_{cd}^{LR}({\phi},{\pi}) \big\}
			\label{piz-phi-amp}\bigg\},
		\end{aligned}
	\end{equation}
	\begin{equation}
		\begin{aligned}
			{ A}(h)=	{ A}(\bar{B}^{0}_d{\to}(\omega-\rho^0\rightarrow \pi ^+\pi ^- ) \pi^{0})=&
			\sum_{\lambda =0,\pm 1} \frac{G_{F}P_{\bar{B}^{0}}\cdot \epsilon^*\left( \lambda \right)\ g^{\rho^0\rightarrow \pi ^+\pi ^- }\epsilon \left( \lambda \right) \cdot \left( p_{\pi ^+}-p_{\pi ^-} \right)}{2\sqrt{2}({{s-m_{\omega}^{2}+im_{\omega} \varGamma _{\omega}})}}\overset{\sim}{\Pi}_{\mathrm{\rho^0\omega}}
			\\
			\times & \bigg\{V_{ub}\,V_{ud}^{\ast}\,
			\big\{ a_{2}\, \big[
			{\cal A}_{ab}^{LL}({\pi},{\omega})
			-{\cal A}_{ab}^{LL}({\omega},{\pi})
			+{\cal A}_{ef}^{LL}({\pi},{\omega})
			+{\cal A}_{ef}^{LL}({\omega},{\pi}) \big]
			\\+ & 
			C_{1}\, \big[
			{\cal A}_{cd}^{LL}({\pi},{\omega})
			-{\cal A}_{cd}^{LL}({\omega},{\pi})
			+{\cal A}_{gh}^{LL}({\pi},{\omega})
			+{\cal A}_{gh}^{LL}({\omega},{\pi}) \big] \big\}
			\\-&
			V_{tb}\,V_{td}^{\ast}\, \big\{
			-( 2\,a_{3}+a_{4}+2\,a_{5}+\frac{1}{2}\,a_{7}
			+ \frac{1}{2}\,a_{9} -\frac{1}{2}\,a_{10} )\,
			{\cal A}_{ab}^{LL}({\omega},{\pi})
			\\-  & 
			( C_{3}+2\,C_{4}-\frac{1}{2}\,C_{9} +\frac{1}{2}\,C_{10} )\,
			{\cal A}_{cd}^{LL}({\omega},{\pi})
			- ( 2\,C_{6}+\frac{1}{2}\,C_{8})\,
			{\cal A}_{cd}^{LR}({\omega},{\pi})
			\\ - & 
			(a_{4}+\frac{3}{2}\,a_{7}-\frac{3}{2}\,a_{9}
			- \frac{1}{2}\,a_{10} )\, \big[
			{\cal A}_{ab}^{LL}({\pi},{\omega})
			+{\cal A}_{ef}^{LL}({\pi},{\omega})
			+{\cal A}_{ef}^{LL}({\omega},{\pi}) \big]
			\\- & 
			(C_{3}- \frac{1}{2}\,C_{9}-\frac{3}{2}\,C_{10} )\, \big[
			{\cal A}_{cd}^{LL}({\pi},{\omega})
			+{\cal A}_{gh}^{LL}({\pi},{\omega})
			+{\cal A}_{gh}^{LL}({\omega},{\pi}) \big]
			\\+ & 
			\frac{3}{2}\,C_{8}\, \big[
			{\cal A}_{cd}^{LR}({\pi},{\omega})
			+{\cal A}_{gh}^{LR}({\pi},{\omega})
			+{\cal A}_{gh}^{LR}({\omega},{\pi}) \big]
			\\-& 
			(a_{6}-\frac{1}{2}\,a_{8})\, \big[
			{\cal A}_{ab}^{SP}({\pi},{\omega})
			+{\cal A}_{ef}^{SP}({\pi},{\omega})
			+{\cal A}_{ef}^{SP}({\omega},{\pi}) \big]
			\\- & 
			( C_{5}-\frac{1}{2}\,C_{7})\, \big[
			{\cal A}_{cd}^{SP}({\pi},{\omega})
			+{\cal A}_{cd}^{SP}({\omega},{\pi})
			+{\cal A}_{gh}^{SP}({\pi},{\omega})
			+{\cal A}_{gh}^{SP}({\omega},{\pi}) \big] \big\}
			\label{piz-w-amp}\bigg\},
		\end{aligned}
	\end{equation}

	\begin{figure}[!htbp]
		\centering
		\begin{minipage}[h]{0.45\textwidth}
			\centering
			
			\includegraphics[height=4cm,width=6.5cm]{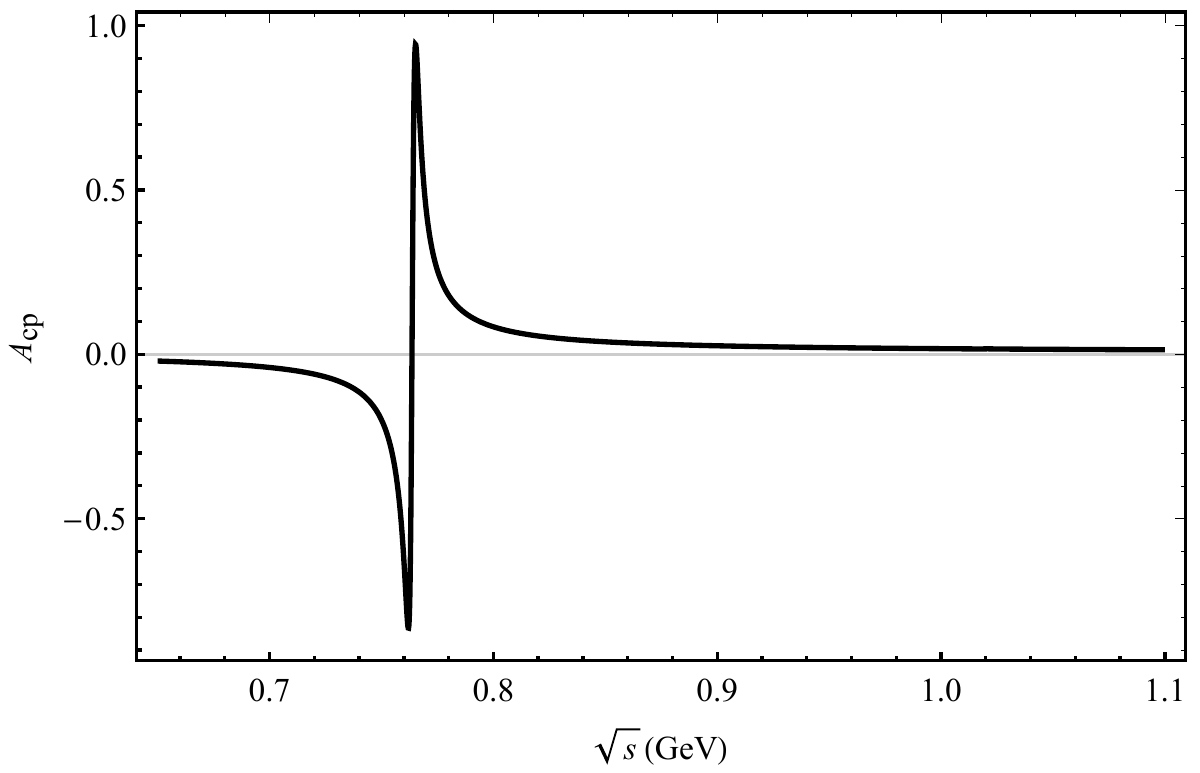}
			\caption{Plot of  $A_{CP}$ as a function of $\sqrt{s}$ corresponding to central parameter values of CKM matrix elements
				for the decay channel of $\bar{B}_{d}^{0}\rightarrow \pi ^+\pi ^-\pi ^0$.}
			\label{fig4}
			\label{fig2}
		\end{minipage}
		\begin{minipage}[h]{0.45\textwidth}
			\centering
			\includegraphics[height=4cm,width=6.5cm]{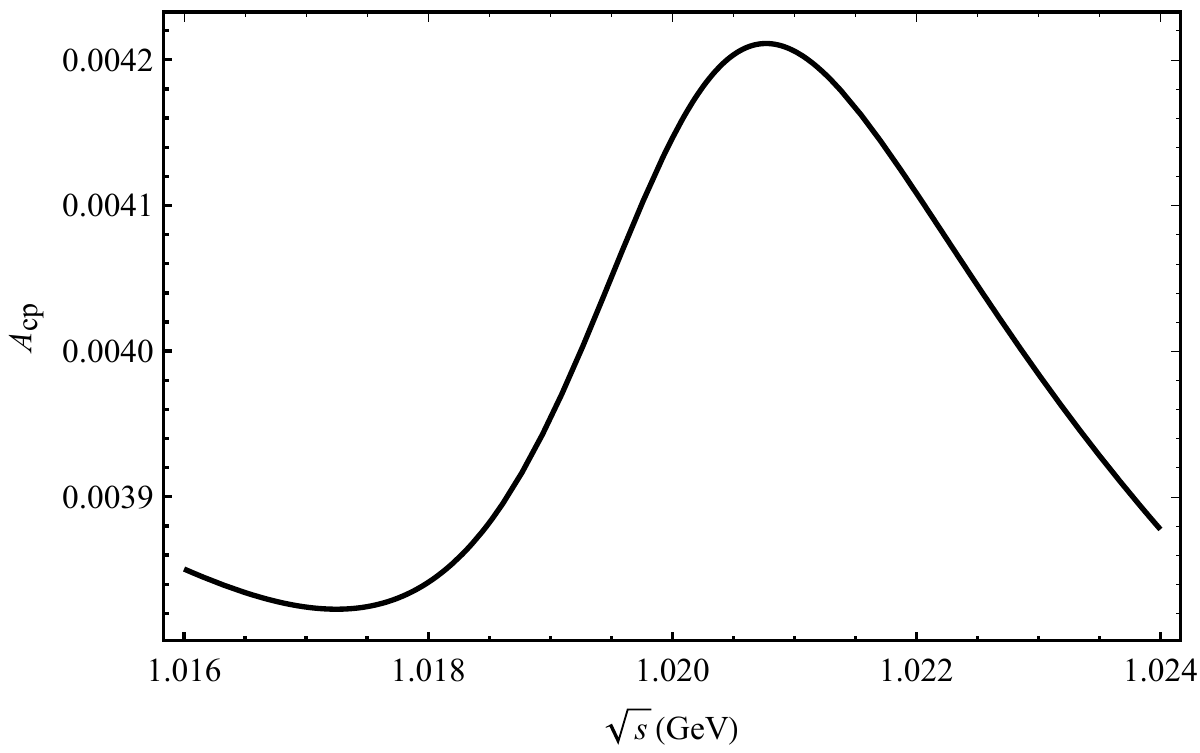}
			\caption{CP asymmetry plot near the invariant mass of $\phi$ during the decay channel of  $ {B}_d^{0} \rightarrow \pi^{+}\pi^{-}\pi^{0}$
				from the central parameter values of CKM matrix elements.}		
			\label{fig6}
		\end{minipage}
	\end{figure}

	In Fig. 5, we present the relationship between the invariant mass of the $\pi^+\pi^{-}$ pair and the CP asymmetry in the decay process $\bar{B}_{d}^{0}\rightarrow \pi ^+\pi ^-\pi ^0$.	
	Within a specific range of invariant masses, we observe a significant variation in CP asymmetry within the $ \rho^{0}-\omega-\phi$ masses region. The maximum value of CP asymmetry reaches $91\%$. Notably, slight peaks are observed at positions corresponding to the mass of the $\phi$ meson from Fig. 6.
	
	To gain further insights into these observations, we investigate the impact of both three-particle and two-particle mixing effects on CP asymmetry in this decay process. By conducting regional integration analysis of CP and examining resonance effects on CP asymmetry, we enhance our understanding of the contributions from different factors to variations in CP violation. Detailed discussions regarding this topic are presented in the subsequent section.

	\subsection{ CP asymmetry analysis of the  $\bar B_d^{0}\rightarrow (\rho^0,\omega,\phi \rightarrow \pi^{+}\pi^{-})\bar K^{0}$  decay process  }

	The above amplitude analysis investigates the impact of vector meson resonance effects on decay processes. During the decay process, vector meson resonances can induce the generation of strong phases. Different vector meson resonances can be observed to have varying effects on the CP asymmetry of decay processes.
	Investigating the resonance effect of vector mesons not only yields valuable physical insights into intermediate mesons in multi-body decay processes, but also presents a novel avenue for future experimental inquiries into CP asymmetry. Currently, there is no experimental data available on the  $\bar B_d^{0}\rightarrow (\rho^0,\omega,\phi \rightarrow \pi^{+}\pi^{-})\bar K^{0}$ decay process.		
	 We present the amplitude formulation of the $\bar B_{d}^{0}\rightarrow (\rho^0,\omega,\phi \rightarrow \pi^{+}\pi^{-})\bar K^{0}$ decay process within the framework of PQCD. 	
	 The three-body amplitude  corresponding to diagram (a) of Fig. 1 can be expressed as follows:
\begin{equation}
	\begin{aligned}
		{ A}(a)=	{ A}(\overline{B}_{d}^{0}{\to}(\rho ^0\rightarrow \pi ^+\pi ^-)\overline{K}^{0})
		&
		=\sum_{\lambda =0,\pm 1} \frac{G_{F}P_{\bar{B}^{0}}\cdot \epsilon^*\left( \lambda \right) g^{\rho ^0\rightarrow \pi ^+\pi ^- }\epsilon \left( \lambda \right) \cdot \left( p_{\pi ^+}-p_{\pi ^-} \right)}{2({{s-m_{\rho ^0}^{2}+im_{\rho ^0} \varGamma _{\rho ^0}}})}
			\\&	
		\times \bigg\{V_{ub}\,V_{us}^{\ast}\,
		\big\{ a_{2}\, \big[
		{\cal A}_{ab}^{LL}({\rho},\overline{K}) \big]
		+ C_{1}\, \big[
		{\cal A}_{cd}^{LL}({\rho},\overline{K})
		\big] \big\}
		+\{V_{tb}\,V_{ts}^{\ast}\, 
		\big\{(a_{4}
		\\ &
		-\frac{1}{2}\,a_{10})\, \big[
		{\cal A}_{ab}^{LL}(\overline{K},{\rho})
		+ {\cal A}_{ef}^{LL}(\overline{K},{\rho}) \big]
		+ (C_{3}
		-\frac{1}{2}\,C_{9})\, \big[
		{\cal A}_{cd}^{LL}(\overline{K},{\rho})	
			\\ & 
		+ {\cal A}_{gh}^{LL}(\overline{K},{\rho}) \big]
		+ (a_{6}-\frac{1}{2}\,a_{8})\, \big[
		{\cal A}_{ab}^{SP}(\overline{K},{\rho})
		+ {\cal A}_{ef}^{SP}(\overline{K},{\rho}) \big]
				\\ & 
		+ (C_{5}-
		\frac{1}{2}\,C_{7})\, \big[
		{\cal A}_{cd}^{SP}(\overline{K},{\rho})
		+ {\cal A}_{gh}^{SP}(\overline{K},{\rho}) \big]
		- \frac{3}{2}\, (a_{7}
		+a_{9})\,
		{\cal A}_{ab}^{LL}({\rho},\overline{K})
		\\ & 
		- \frac{3}{2}\,C_{8}\,
		{\cal A}_{cd}^{LR}({\rho},\overline{K})
		- \frac{3}{2}\, C_{10}\,
		{\cal A}_{cd}^{LL}({\rho},\overline{K})  \big\}
		\label{kz-rhoz-amp}\bigg\}.
	\end{aligned}
\end{equation}
 In Fig. 1, diagram (h) depicts the decay of $\rho^0$ meson into the pair of $\pi^{+}\pi^{-}$ mesons via the $\omega$ resonance, indicated by a black dot. Diagram (e) is similar to diagram (h), except for the decay of $\rho^0$ meson through the $\phi$ resonance.
 The amplitudes associated with diagram (h)  and (e) of Fig. 1 can be expressed as follows:
 
	\begin{equation}
	\begin{aligned}
		{ A}(h)=	{A}(\overline{B}_{d}^{0}{\to}(\omega-\rho^0\rightarrow \pi ^+\pi ^-) \overline{K}^{0})
		&=\sum_{\lambda =0,\pm 1} \frac{G_{F}P_{\bar{B}^{0}}\cdot \epsilon^*\left( \lambda \right)\ g^{\rho^0\rightarrow \pi ^+\pi ^- }\epsilon \left( \lambda \right) \cdot \left( p_{\pi ^+}-p_{\pi ^-} \right)}{2(s-m_{\rho^0}^{2}+im_{\rho^0} \varGamma _{\rho^0})({s-m_{\omega}^{2}+im_{\omega} \varGamma_{\omega}})}\overset{\sim}{\Pi}_{\mathrm{\rho^0\omega}}
		\\&
		\times \bigg\{V_{ub}\,V_{us}^{\ast}\,
		\big\{ a_{2}\, \big[
		{\cal A}_{ab}^{LL}({\omega},\overline{K}) \big]
		+ C_{1}\, \big[
		{\cal A}_{cd}^{LL}({\omega},\overline{K})
		\big] \big\}
		\\&
		-\{V_{tb}\,V_{ts}^{\ast}\, \big\{
		(a_{4}-\frac{1}{2}\,a_{10})\, \big[
		{\cal A}_{ab}^{LL}(\overline{K},{\omega})
		+ {\cal A}_{ef}^{LL}(\overline{K},{\omega}) \big]	
		+ (C_{3}	\\ &
		-\frac{1}{2}\,C_{9})\, \big[
		{\cal A}_{cd}^{LL}(\overline{K},{\omega})
		+ {\cal A}_{gh}^{LL}(\overline{K},{\omega}) \big]
		+ (a_{6}-\frac{1}{2}\,a_{8})\, \big[
		{\cal A}_{ab}^{SP}(\overline{K},{\omega})
		\\ &  
		+ {\cal A}_{ef}^{SP}(\overline{K},{\omega}) \big]
		+ (C_{5}-\frac{1}{2}\,C_{7})\, \big[
		{\cal A}_{cd}^{SP}(\overline{K},{\omega})
		+ {\cal A}_{gh}^{SP}(\overline{K},{\omega}) \big]
		\\ & 
		+ (2\,a_{3}+2\,a_{5}+\frac{1}{2}\,a_{7}
		+\frac{1}{2}\,a_{9})\,
		{\cal A}_{ab}^{LL}({\omega},\overline{K}) 
		+ (2\,C_{4} \\ & 
		+\frac{1}{2}\,C_{10})\,
		{\cal A}_{cd}^{LL}({\omega},\overline{K})
		+ (2\,C_{6}+\frac{1}{2}\,C_{8})\,
		{\cal A}_{cd}^{LR}({\omega},\overline{K}) \big\}
		\label{kz-w-amp}\bigg\},
	\end{aligned}
\end{equation}


\begin{equation}
	\begin{aligned}
		{ A}(e)=	{ A}(\overline{B}_{d}^{0}{\to}(\phi-\rho^0\rightarrow \pi ^+\pi ^- )\overline{K}^{0})
		 &=
		\sum_{\lambda =0,\pm 1} \frac{-G_{F}P_{\bar{B}^{0}}\cdot \epsilon^*\left( \lambda \right)\ g^{\rho^0\rightarrow \pi ^+\pi ^- }\epsilon \left( \lambda \right) \cdot \left( p_{\pi ^+}-p_{\pi ^-} \right)}{\sqrt{2}({{s-m_{\phi}^{2}+im_{\phi} \varGamma _{\phi}})}}\overset{\sim}{\Pi}_{\mathrm{\rho\phi}}
		\\&
		\times\bigg\{ V_{tb}\,V_{ts}^{\ast}\, \big\{
		(a_{3}+a_{4}+a_{5}-\frac{1}{2}\,a_{7}
		-\frac{1}{2}\,a_{9}-\frac{1}{2}\,a_{10})\,
		{\cal A}_{ab}^{LL}({\phi},\overline{K})
		\\ & 
		+(a_{4}-\frac{1}{2}\,a_{10})\,
		{\cal A}_{ef}^{LL}({\phi},\overline{K})
		+(a_{6}-\frac{1}{2}\,a_{8})\,
		{\cal A}_{ef}^{SP}({\phi},\overline{K})
		\\ & 
		+ (C_{3}+C_{4}-\frac{1}{2}\,C_{9}-\frac{1}{2}\,C_{10})\,
		{\cal A}_{cd}^{LL}({\phi},\overline{K})
		+ (C_{6}-\frac{1}{2}\,C_{8})\,
		{\cal A}_{cd}^{LR}({\phi},\overline{K})		\\ & 
	+ (C_{5}-\frac{1}{2}\,C_{7})\, \big[
{\cal A}_{cd}^{SP}({\phi},\overline{K})
+{\cal A}_{gh}^{SP}({\phi},\overline{K}) \big]
+ (C_{3}-\frac{1}{2}\,C_{9})\,
{\cal A}_{gh}^{LL}({\phi},\overline{K}) \big\}
\label{kz-phi-amp}\bigg\}.
	\end{aligned}
\end{equation}

	\begin{figure}[!htbp]
		\centering
		\begin{minipage}[h]{1\textwidth}
			\centering
			\includegraphics[height=6cm,width=9cm]{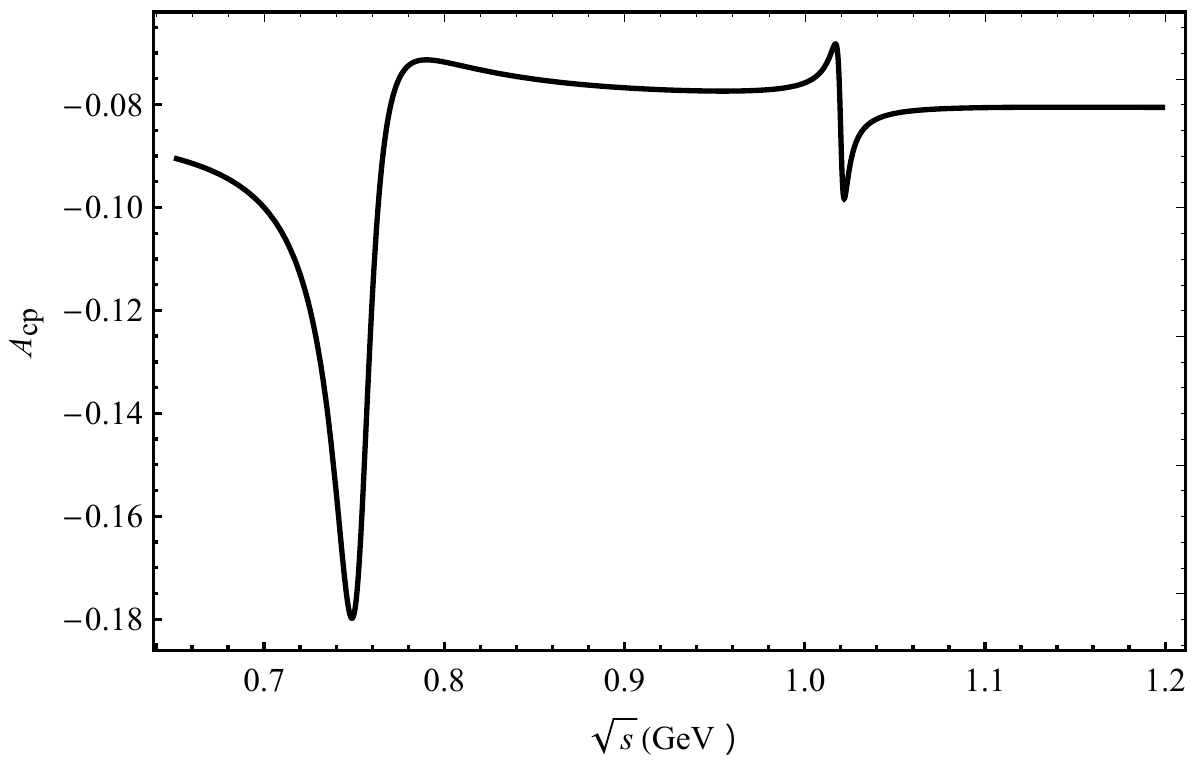}
			\caption{Plot of  $A_{CP}$ as a function of $\sqrt{s}$ corresponding to central parameter values of CKM matrix elements for the decay channel of $\bar{B}_{d}^{0}\rightarrow \pi ^+\pi ^-\bar K^{0}$.}
			\label{fig5}
		\end{minipage}
	\end{figure}
	The results of CP asymmetry for the $\bar{B}_{d}^{0}\rightarrow \pi ^+\pi ^-\bar K^{0}$ decay process is presented in Fig. 7. As depicted in Fig. 7, it is evident that the CP asymmetry of the $\bar{B}_{d}^{0}\rightarrow \pi ^+\pi ^-\bar K^{0}$ channel undergoes variations when the invariant masses of the pair of $\pi^+\pi^{-}$ mesons encompass the ranges corresponding to resonances such as $\omega$ and $\phi$. Notably, a maximum CP asymmetry of $-18\%$ can be achieved.
	
	The CP asymmetry in the decay process $\bar{B}_{d}^{0}\rightarrow \pi ^+\pi ^-\bar K^{0}$ exhibits a significant change when the invariant mass of the pair of $\pi^+\pi^{-}$ mesons approaches 0.75 GeV, reaching a peak value of $-18\%$. This behavior can be attributed to the interference effect arising from the $\rho-\omega$ mixing mechanism, which is expected to occur within the range of 0.7 GeV - 0.8 GeV. Additionally, small peaks are observed in the invariant mass range corresponding to $\phi$. 	
In the decay of ${B}_{d}^{0}{\to}(\phi\rightarrow \pi^{+}\pi^{-}) \bar{K}^{0}$, penguin graph contributions play a significant role. Notably, the resonance effect arising from $\rho^0-\phi$ mixing exerts a substantial influence on CP asymmetry, resulting in an associated value of $-9\%$.

	\section{\label{cal}Analysis of regional CP asymmetry in the decay process}
	The narrow width approximation (NWA) is employed in our calculation to decompose the three body decay process into two successive quasi-two-body decays: ${B}\rightarrow (R \rightarrow\pi^{+} \pi^{-})P$, where R denotes the resonance state and P represents either $\pi$ or K meson.
Considering the intermediate resonance state R, we introduce the correction factor $\eta_{R}$ to define the expression:
\begin{equation}
	\begin{aligned}
		\eta_{R}=\frac{\mathcal{B} \left( B\rightarrow RP_3\rightarrow P_1P_2P_3 \right) }{\mathcal{B} \left( B\rightarrow RP_3 \right) \mathcal{B} \left( B\rightarrow P_1P2 \right)},
	\end{aligned}
\end{equation}
where shows the relationship  between the branch ratio measured around the resonance region and the branch ratio of the three-body decay of the B meson \cite{Chua:2022owm}. According to the QCDF method, the correction factor $\eta_{R}$ is about $7\%$ during the decay of  ${B}_u^{-}\rightarrow  \pi^{+}\pi^{-}\pi^{-}$ \cite{Cheng:2022ysn,Cheng:2020mna}. 
When calculating CP asymmetry, both the numerator and denominator contain the correction factor $\eta_{R}$, which can cancel each other. 

We investigate the resonant and non-resonant contributions within a specific region, examining how various particle resonances contribute to CP asymmetry. Our calculations involve integrating the differential CP asymmetry parameter in both the numerator and denominator simultaneously, allowing us to derive regional CP asymmetry.

	\begin{table}[h]
		
		{\renewcommand
			\scalebox{1}
			\renewcommand{\arraystretch}{2} 
			\setlength{\tabcolsep}{4mm}
			\begin{center}
				\caption{The peak regional $(0.70GeV\leq \sqrt{s}\leq1.10GeV)$ integral of  $\mathrm{A}^{\Omega} _{\mathrm{CP}}$ from different resonance rangs for  $\bar B_{d} \rightarrow  \pi^{+}\pi^{-} K^{0}$, $\bar B_{d} \rightarrow  \pi^{+}\pi^{-}  \pi^{0}$ and  $\ B_{u}^- \rightarrow  \pi^{+}\pi^{-} K^{-}$, $\ B_{u}^- \rightarrow  \pi^{+}\pi^{-} \pi^{-}$  decay processes.}	
				\begin{tabular}{lllllllll}
					\hline
					\hline\
					Decay channel      &This work &Previous measurements(no mixing)   \\ \hline		
					{\makecell[c]{$\bar B_{d}^0 \rightarrow  \pi^{+}\pi^{-} \bar K^{0}$}}
					&$\mathrm{-0.1683\pm 0.0013\pm0.0000}$ ($\rho^0$)
					&  \\
					&$\mathrm{-0.0847\pm 0.0000\pm0.0084}$ ($\rho^0-\omega$ mixing )
					& -----------------------------\\
					&$\mathrm{-0.0813\pm 0.0064\pm0.0052}$ ($\rho^0-\phi$  mixing)
					& \\
					&$\mathrm{-0.0847\pm 0.0056\pm0.0089}$ ($\phi-\rho^0-\omega$ mixing )
					& \\
					{\makecell[c]{ $\bar B_{d}^0 \rightarrow  \pi^{+}\pi^{-}  \pi^{0}$ }}
					& $\mathrm{-0.0054\pm 0.0004\pm0.0011}$ ($\rho^0$)
					& \\
					& $\mathrm{0.0173\pm 0.0109\pm0.0015}$ 	($\rho^0-\omega$ mixing )
					&-----------------------------\\
					&$\mathrm{-0.0055\pm 0.0006\pm0.0011} $ ($\rho^0-\phi$ mixing )
					& \\	
					&$\mathrm{0.0147\pm 0.0014\pm0.0086} $ ($\phi-\rho^0-\omega$  mixing)
					& \\	
					
					{\makecell[c]{$\ B_{u}^- \rightarrow  \pi^{+}\pi^{-} K^{-}$}}		
					& $\mathrm{0.2093\pm0.0206\pm0.0044}$ ($\rho^0$)	
					& 0.150$\pm$0.019$\pm$0.011 LHCb\cite{LHCb:2022tuk} \\
					& $\mathrm{0.1771\pm0.0084\pm0.0061}$ 	($\rho^0-\omega$ mixing )
					&0.44$\pm$0.10$\pm$0.04 BaBar\cite{BaBar:2008lpx} \\
					& $\mathrm{0.2109\pm0.0207\pm0.0023}$ 	 ($\rho^0-\phi$  mixing)
					&0.30$\pm$0.11$\pm$0.02 Belle\cite{Belle:2005rpz} \\
					& $\mathrm{0.3470\pm0.0310\pm0.0709}$ 	($\phi-\rho^0-\omega$  mixing)
					&  \\
					
					{\makecell[c]{$\ B_{u}^- \rightarrow  \pi^{+}\pi^{-} \pi^{-}$}} 
					
					&$ \mathrm{0.0065\pm 0.0014\pm0.0031}$($\rho^0$)
					&-0.004$\pm$0.017$\pm$0.009 LHCb\cite{LHCb:2022tuk}\\
					& $\mathrm{0.0256\pm0.0013\pm0.0016}$ 		($\rho^0-\omega$  mixing)
					&0.30$\pm$0.11$\pm$0.02 Belle\cite{Belle:2005rpz} \\
					& $\mathrm{0.0076\pm0.0006\pm0.0023}$  ($\rho^0-\phi$  mixing)
					&  \\
					&$\mathrm{0.0260\pm 0.0034\pm0.0047}$  ($\phi-\rho^0-\omega$ mixing)
					&  \\		
					\hline
					\hline
				\end{tabular}
				\label{tab:syst_uncert}
		\end{center}}
	\end{table}

In the Table II, we present a comparison between our calculated results and the existing experimental data. Notably, intriguing observations of significant regional CP asymmetry have been reported in experiments conducted by LHCb, BaBar and Belle collaborations. Subsequently, we investigate the influence of different resonance effects on regional CP asymmetry. Due to the close masses of $\rho^0$ and $\omega$ mesons in these experiments, distinguishing between them becomes challenging. To address this issue, we incorporate this scenario into our calculations using the PQCD method within the resonance framework for decay channels such as $\bar B_{d}^0 \rightarrow  \pi^{+}\pi^{-} \bar K^{0}$, $\bar B_{d}^0 \rightarrow  \pi^{+}\pi^{-}  \pi^{0}$, $B_{u}^- \rightarrow  \pi^{+}\pi^{-} K^{-}$ and $B_{u}^- \rightarrow  \pi^{+}\pi^{-} \pi^{-}$. Our results align with experimental error ranges, validating the accuracy of our calculation method. Specifically in the decay channel of $\bar B_{d}^0 \rightarrow  \pi^{+}\pi^{-} \bar K^{0}$ , we observe that resonance contributions from three-particle mixtures significantly impact CP asymmetry.

In the decay process of ${\bar B_{d}^0 \rightarrow  \pi^{+}\pi^{-} \bar K^{0}}$ , a significant CP asymmetry value can be obtained by the intermediate  $\rho^{0}$ meson. 
Considering the resonance effect of vector mesons, the CP asymmetry value of three-body decay has changed significantly. 
Combining the data in the Table II, the resonance effect has played a role in suppressing the CP asymmetry value of three-body decay. The suppression effect of two vector meson and three vector meson mixture resonance on the CP asymmetry value of the decay process is not significantly different.

In the decay process of ${\bar B_{d}^0 \rightarrow  \pi^{+}\pi^{-}  \pi^{0}}$, the mixed resonance effect of $\rho^{0}-\omega$ and $\phi-\rho^0-\omega$ has made a substantial contribution to CP asymmetry. 
 The CP asymmetry value associated with the direct decay of the $\rho^{0}$ meson closely approximates the value obtained under the $\rho^0-\phi$ mixed resonance mechanism, indicating that this process is minimally affected by the presence of $\rho^0-\phi$ mixed resonance.
 Currently, there is a lack of experimental studies investigating the CP asymmetry in the ${\bar{B}_{d}^0 \rightarrow\pi^{+}\pi^{-} K(\pi)}$ three-body decay process. We anticipate that our prediction can serve as a valuable reference for future experiments.

 The CP asymmetry value resulting from the intermediate $\rho^{0}$ meson in the ${B_{u} \rightarrow\pi^{+}\pi^{-} K(\pi)}$ decay process falls within the experimental error range. 
 The decay process {$B_{u}^- \rightarrow  \pi^{+}\pi^{-} K^{-}$} exhibits a suppression effect due to the presence of $\rho^0-\omega$ resonance, while significant contributions to CP asymmetry are observed from both $\rho^0-\omega$ and $\phi-\rho^0-\omega$ resonances. Notably, the three-particle resonance contribution surpasses that of the $\rho^0-\phi$ mixed resonance in terms of CP asymmetry.	
	 In the decay process of $B_u^- \rightarrow \pi^+\pi^-\pi^-$, various mixed resonances have been considered in this study, leading to an overall enhancement in the CP asymmetry value. The contributions from the $\rho^0-\omega$ and $\phi-\rho^0-\omega$ mixed resonances are approximately equal in terms of CP asymmetry, while the mixing mechanism involving $\rho^0-\phi$ also contributes to CP asymmetry but with a smaller effect.

	For vector particle resonances, the decay processes $\bar B_{d} \rightarrow  \pi^{+}\pi^{-}  \pi^{0}$ and $B_{u}^- \rightarrow  \pi^{+}\pi^{-} \pi^{-}$ exhibit significant contributions to CP asymmetry compared to non-resonant decays. It is noteworthy that clear manifestations of CP asymmetry are observed within the resonance range. In the decay channel of $B_{u}^- \rightarrow  \pi^{+}\pi^{-} K^{-}$, resonance effects contribute more prominently to CP asymmetry, resulting in increased regional values similar to those observed in the decay channel $\bar B_{d} \rightarrow  \pi^{+}\pi^{-}K^0$ within their respective invariant mass ranges.

	The perturbative calculation of the $1/m_b$ power correction is subject to uncertainties. Uncertainties in the results can arise from data errors, including variations in CKM parameters and other parameters such as form factors, decay constants and B-meson wave functions that involve distributed amplitudes in vector meson wave functions.	
	By utilizing the central values of these parameters, we initially calculate numerical outcomes for CP asymmetry and incorporate standard deviations to establish the error range presented in Table II. The effect of data error on regional CP asymmetry is negligible; thus, it can be disregarded without further deliberation.

	\section{\label{sum}SUMMARY AND DISCUSSION}
	
	We conduct a comprehensive analysis on the CP asymmetry in the three-body decay of B meson, focusing specifically regions on the invariant mass of the pair of $\pi^{+}\pi^{-}$ mesons. 	
	The research findings revealed an intriguing phenomenon: there are significant changes in CP asymmetry in different resonance regions (e.g., resonances of $\rho^0$, $\omega$ and $\phi$ mesons).
	This finding suggests that these resonances play a significant role in influencing the decay dynamics and subsequent CP asymmetry.	
	The presence of such distinct changes in CP asymmetry across different resonance regions provides valuable insights into our understanding of fundamental particle interactions. It highlights how specific energy regimes can impact particle decays and their associated symmetries.

	We quantify regional CP asymmetry by integrating over the phase space. In decays such as $\bar B_{d}^0 \rightarrow  \pi^{+}\pi^{-}\bar K^{0}$, $\bar B_{d}^0 \rightarrow  \pi^{+}\pi^{-}  \pi^{0}$, $B_{u}^- \rightarrow  \pi^{+}\pi^{-} K^{-}$ and $B_{u}^- \rightarrow  \pi^{+}\pi^{-} \pi^{-}$, we observe CP asymmetry arising from contributions of two-meson mixing and three-meson mixing processes. Notably, significant regional CP asymmetry is observed when involving $\rho^0-\phi- \omega$ mixing. Experimental detection of regional CP asymmetry can be achieved by reconstructing resonant states of $\rho^0$ , $\omega$ and $\phi$ mesons within their respective resonance regions.

	Recently, the LHCb experimental group has made significant progress in investigating the three-body decay of B meson and has obtained noteworthy findings. By analyzing previous experimental data, they have measured direct CP asymmetry in various decay modes such as $B^{\pm} \rightarrow K^{+} K^{-} K^\pm$, $B^{\pm} \rightarrow \pi^{+} \pi^{-} K^\pm$, $B^{\pm} \rightarrow \pi^{+} \pi^{-} \pi^\pm$ and $B^{\pm} \rightarrow K^{+} K^{-}\pi^\pm$. Building upon the achievements of LHCb experiments, future investigations are expected to primarily focus on exploring regional CP asymmetry of the three-body decays of $B$ meson 
	at the resonance regions of $\rho^0$, $\omega$ and $\phi$ mesons.

	\section*{Acknowledgements}
	This work was supported by  Natural Science Foundation of Henan (Project No. 232300420115) and National Natural Science Foundation of China (Project No.12275024).
	
	
	\end{spacing}
\end{document}